%% file: main.tex
\documentclass[10pt,journal]{IEEEtran}
\IEEEoverridecommandlockouts 
\input{setting-ieee}
\graphicspath{{}{figs/}}

\usepackage[ruled,vlined, algo2e, linesnumbered]{algorithm2e}
\Crefname{algocf}{Algorithm}{Algorithms} 

\usepackage{textcomp}
\usepackage{amsmath}
\usepackage{bm} 
\usepackage{subfigure}
\usepackage{diagbox}
\newcommand{\RNum}[1]{\uppercase\expandafter{\romannumeral #1\relax}}

\begin{document}

\title{
    CUTh-Solver: GPU-Accelerated Sparse Matrix Solver for High-Resolution Thermal Simulation of 3D ICs
}

\author{
    Chenghan Wang, \quad
    Zhen Zhuang, \quad
    Shui Jiang, \quad
    Siyuan Liang, \quad
    Xiaoman Yang, \quad
    Kai Zhu, \quad
    Darong Huang, \quad
    Luis Costero, \quad
    Rongmei Chen, \quad
    Tsung-Wei Huang, \quad
    David Atienza, \quad
    Tsung-Yi Ho

    
    \thanks{Chenghan Wang, Zhen Zhuang, Shui Jiang, Siyuan Liang, Xiaoman Yang and Tsung-Yi Ho are with the Department of Computer Science and Engineering, The Chinese University of Hong Kong, NT, Hong Kong SAR. (\textit{Corresponding author: Zhen Zhuang})}
    \thanks{Kai Zhu, Darong Huang, and David Atienza are with Embedded Systems Laboratory (ESL), École Polytechnique Fédérale de Lausanne (EPFL), Lausanne, Switzerland.}
    \thanks{Darong Huang is also with Department of Electrical and Computer Engineering, The University of Hong Kong, Hong Kong Island, Hong Kong SAR.}
    \thanks{Luis Costero is with the Department of Computer Architecture and System Engineering, Universidad Complutense de Madrid (UCM), Madrid, Spain.}
    \thanks{Rongmei Chen is with the School of Electronics, Peking University,
    Beijing, China.}
    \thanks{Tsung-Wei Huang is with the Department of Electrical and Computer Engineering (ECE), the University of Wisconsin at Madison, Wisconsin, USA.}
}

\maketitle
\thispagestyle{plain}
\pagestyle{plain}

\input{chapters/0_abstract}
\input{chapters/1_intro}
\input{chapters/2_relatedworks}
\input{chapters/3_preliminaries}
\input{chapters/4_methodology}
\input{chapters/5_experiments}
\input{chapters/6_conclusion}
\input{chapters/7_acknowledgment}

\bibliographystyle{IEEEtran}
\bibliography{ref/references}

\end{document}

%% file: setting-ieee.tex
\usepackage{blkarray}                                      
\usepackage{graphicx}                                      
\usepackage{amsmath}
\usepackage{amssymb}
\usepackage{amsfonts}
\usepackage{amsthm}
\usepackage[mathcal]{eucal}
\usepackage{mathrsfs}
\usepackage{booktabs}
\usepackage{enumerate}
\usepackage{multirow}
\usepackage{subfigure}
\usepackage{color}
\usepackage{cite}                                          
\usepackage{comment}                                       
\usepackage{soul}                                          
\soulregister\cite7
\soulregister\ref7
\soulregister\pageref7
\usepackage{etoolbox}                                      
\usepackage{url}
\usepackage{nth}                                           
\usepackage{bm}                                            
\usepackage{courier}
\usepackage{balance}
\usepackage{threeparttable}
\usepackage{xcolor,colortbl}
\usepackage{footnote}
\usepackage{listings}
\usepackage{setspace}                                      
\usepackage[inline]{enumitem}
\usepackage{verbatim}
\usepackage[bookmarks=false]{hyperref}
\hypersetup{
    colorlinks = true,
    citecolor  = blue,
    linkcolor  = blue,
    urlcolor   = blue,
}
\usepackage{tikz}
\usetikzlibrary{patterns,snakes}
\usetikzlibrary{positioning,calc,fit,decorations.pathmorphing,shapes.geometric, shapes.gates.logic.US, calc}
\usetikzlibrary{arrows,arrows.meta,decorations.markings,shapes,shapes.arrows}
\usetikzlibrary{decorations,decorations.pathreplacing}
\usetikzlibrary{backgrounds}
\usepackage{pgfplots}
\usepackage{pgfplotstable}
\usepgfplotslibrary{groupplots}
\usepackage{scalefnt}
\pgfplotsset{compat=newest}
\usepackage{caption}
\usepackage{pifont}                                        
\usepackage{cleveref}
\Crefformat{figure}{Fig.~#2#1#3}                           
\Crefname{subfigure}{Fig.}{Figs.}
\Crefname{figure}{Fig.}{Figs.}
\Crefformat{table}{TABLE~#2#1#3}                           
\usepackage[figuresright]{rotating}
\usepackage{algorithm}
\iffalse
\usepackage{algpseudocode}                                 
\algrenewcommand\textproc{\texttt}
\makeatletter
\let\OldStatex\Statex
\renewcommand{\Statex}[1][3]{%
  \setlength\@tempdima{\algorithmicindent}%
  \OldStatex\hskip\dimexpr#1\@tempdima\relax
}
\makeatother
\else
\usepackage{algorithmic}
\fi

\definecolor{CUHKorange}{RGB}{244,106,18} 
\definecolor{CUHKblue}{RGB}{0,111,190}    
\definecolor{CUHKgreen}{RGB}{0,127,128}   
\definecolor{CUHKred}{RGB}{228,46,36}     
\definecolor{CUHKyellow}{RGB}{198,148,34} 
\definecolor{CUHKdark}{RGB}{114,44,114}   
\definecolor{CUHKmiddle}{RGB}{144,44,144} 
\definecolor{CUHKlight}{RGB}{167,44,167} 
\definecolor{CUHKpurple}{RGB}{117,15,109}
\definecolor{CUHKgold}{RGB}{221,163,0}
\definecolor{CUHKribbon}{RGB}{244,223,176}
\definecolor{CUHKblack}{RGB}{34,24,21}

\renewcommand{\bf}[1]{\textbf{#1}}
\renewcommand{\tt}[1]{\texttt{#1}}

\usepackage{tcolorbox}
\tcbuselibrary{skins,breakable}
    {\endtcolorbox}
    {\endtcolorbox}

\usepackage[top=0.80in,bottom=0.88in,left=0.63in,right=0.63in]{geometry}
\crefname{mytheorem}{Theorem}{Theorems}
\crefname{mylemma}{Lemma}{Lemmas}
\crefname{myclaim}{Claim}{Claims}
\crefname{myproperty}{Property}{Properties}
\crefname{mycorollary}{Corollary}{Corollaries}

\RequirePackage[normalem]{ulem} 
\RequirePackage{color}\definecolor{RED}{rgb}{1,0,0}\definecolor{BLUE}{rgb}{0,0,1} 

%% file: chapters/0_abstract.tex
\begin{abstract}

Coarse-grained thermal simulation tends to underestimate localized thermal issues, potentially missing critical hotspots. Accurate analysis, therefore, demands fine-grained information, which dramatically increases grid resolution and thus computational workload. Fortunately, the coefficient matrices are often sparse with regular sparsity patterns, offering optimization opportunities. However, existing general-purpose matrix solvers on GPUs rarely exploit these domain-specific properties, thereby encountering bottlenecks in data storage, memory access, parallelism, computational efficiency, and hardware utilization. Therefore, we propose \textbf{CUTh-Solver}, a co-designed GPU-accelerated Preconditioned Conjugate Gradient (PCG)-based sparse solver framework for Symmetric Positive Definite (SPD) systems arising from high-resolution steady-state and transient 3D IC thermal simulation. For data storage, CUTh-Solver condenses the Diagonal (DIA) storage format to remove redundancy. To optimize the memory access, CUTh-Solver employs diagonal-wise SpMV to achieve coalesced memory access.
We further observe a critical conflict between parallelism and preconditioning quality and thus adopt a high-parallelism preconditioning strategy. To improve computational efficiency and hardware utilization, we employ an adaptive fine-grained mixed-precision strategy that leverages diverse floating-point units to avoid resource contention, enhancing throughput without compromising numerical stability. Experimental results show that CUTh-Solver achieves up to $25.8\times$ speedup over GPU-accelerated COMSOL Multiphysics 6.4 and over $3\times$ speedup over NVIDIA's native general-purpose libraries (AmgX, cuSPARSE, cuDSS). Ablation studies validate the individual contribution of each optimization. \textcolor{black}{The code is available at: https://github.com/Chenghan-Wang/CUTh-Solver.}

\end{abstract}

%% file: chapters/1_intro.tex
\section{Introduction}
\label{sec:1}

As a pivotal solution to transcend the scaling limits of Moore's Law, 3D integration offers substantial advantages in integration density, interconnect latency, and energy efficiency~\cite{lim_3dic}. However, vertical stacking of active components significantly exacerbates thermal density, posing severe challenges to thermal management~\cite{imec_thermal_1, imec_thermal_2, imec_thermal_3}.

\input{figs/intro_story.tex}

Accurate thermal simulation requires fine-grained design features extracted from IC design files~\cite{parameter_extraction_1, parameter_extraction_2, parameter_extraction_4}. Coarse-grained thermal analysis risks obscuring local hotspots by averaging out localized peaks~\cite{PACT, multiscale_iedm}. As illustrated in \Cref{intro_story}(a), the maximum power density at a resolution of 1-\textmu m can be up to $6.3\times$ higher than at a resolution of 20-\textmu m~\cite{multiscale_iedm}, leading to temperature discrepancies sometimes exceeding 10~\textcelsius. The latest 3D-ICE 4.0 also demonstrates that coarse-grained modeling can lead to a significant underestimate of temperature\cite{3dice4.0}. However, meshing millimeter-scale 3D IC packaging at micron-level granularity produces high-resolution simulation cells~\cite{multiscale_iedm}.

With high-resolution simulation cells, a sparse large-scale linear system $\bm{Ax=b}$ is obtained, whose numerical solution constitutes the main computational bottleneck, often exceeding $60\%$ of the overall simulation runtime~\cite{nvidia_amgx, taco}. Consequently, the design and implementation of an efficient sparse solver is critical to enhancing simulation throughput.

Sparse linear solvers fall into direct and iterative methods~\cite{solver_survey}. Direct solvers (e.g., Cholesky decomposition) are numerically robust but suffer from prohibitive runtime and memory overheads due to fill-in effects, as shown in \Cref{intro_story}(d), making them ill-suited for high-resolution thermal simulation~\cite{jinzhou_2024_dac}. Despite this, many popular open-source thermal tools, including HotSpot 7.0~\cite{hotspot_7.0}, 3D-ICE 4.0~\cite{3dice4.0}, PACT~\cite{PACT}, and MFIT~\cite{MFIT}, still rely on CPU-based direct solvers, which inherently limit their scalability. Iterative methods offer superior efficiency in both throughput and memory footprint~\cite{Yu_2024_TCAD, siyuan_powerplane}. In particular, the Preconditioned Conjugate Gradient (PCG) method has become the de facto standard for large-scale Symmetric Positive Definite (SPD) systems~\cite{2001_DAC_ICCPCG, Yu_2021_ICCAD}. Nevertheless, as shown in \Cref{intro_story}(d), PCG solvers on CPU platforms still require hundreds of seconds for large-scale linear systems with tens of millions of unknowns.

The limited scalability of CPU-based sparse solvers drives a shift toward GPU acceleration. A notable example is that the latest COMSOL Multiphysics 6.4 ( released in November 2025) integrates NVIDIA cuDSS to accelerate direct solvers across its diverse physics modules~\cite{comsol_gpu_acceleration}. However, existing GPU-accelerated general-purpose sparse libraries, including cuDSS, struggle to fully leverage the domain-specific knowledge of 3D IC heat transfer, facing fundamental bottlenecks in data storage, GPU memory access, parallelism, computational efficiency, and hardware utilization:

\textbf{Data Storage:} Matrices from solid-state 3D IC thermal simulation are typically banded and SPD. However, general-purpose storage formats do not use symmetry by default. While the DIA format leverages the band property, it introduces excessive zero-padding when the diagonal bandwidth is large, which is a common scenario in high-resolution 3D IC thermal simulation. Therefore, with these domain-specific properties, general-purpose storage formats can be condensed to further improve the efficiency of data storage.

\textbf{GPU Memory Access:} General-purpose sparse formats often incur indirect addressing and irregular memory access in SpMV operations, causing frequent high-latency global DRAM access. Leveraging the regular sparsity of matrices arising from 3D IC thermal analysis, coalesced memory access can be achieved in SpMV operations to lower the frequency of the costly global DRAM access.

\textbf{Parallelism:} Preconditioners suffer from the conflict between convergence rate and parallelism. CPU-favored preconditioners involve complex data dependencies that are difficult to parallelize. We observe that after migration to GPUs, their limited parallelism can degrade overall performance, even worse than that of vanilla CG solvers. This observation suggests that GPU acceleration demands a different preconditioning strategy from CPUs, one that balances mathematical convergence with hardware parallelism.

\textbf{Computational Efficiency and Hardware Utilization:} Driven by machine learning workloads, recent consumer-grade GPUs are drastically reducing double-precision (FP64) hardware resources. However, general-purpose scientific computing conventionally defaults to FP64 for numerical stability, causing massive floating-point operations to queue for increasingly scarce FP64 units while abundant single- and half-precision units sit idle. Domain-specific insights from 3D IC thermal simulation can effectively enable lower-precision arithmetic in precision-insensitive steps to avoid resource contention, improving hardware utilization and computational efficiency without sacrificing numerical stability.

Therefore, we present \textbf{CUTh-Solver}, an open-source and GPU-accelerated sparse solver framework specifically tailored for high-resolution steady-state and transient thermal analysis of 3D ICs. CUTh-Solver is designed to bridge the efficiency gap inherent in general-purpose sparse solvers by fully leveraging prior domain-specific insights. CUTh-Solver is publicly available, providing intuitive interfaces and supporting seamless, out-of-the-box integration for both steady-state and transient thermal simulation. The primary contributions of this work are summarized as follows:
\begin{itemize}
    \item \textbf{Redundancy-Free Storage Format:} Leveraging the SPD matrix property, we further condense the conventional DIA format into \textbf{Quad-Diag}, a redundancy-free and GPU-friendly storage format specifically tailored for 7-diagonal matrices arising from high-resolution 3D IC thermal simulation.
    \item \textbf{Coalesced Memory Access in SpMV:} Quad-Diag facilitates a diagonal-wise SpMV computation pattern, in which vector elements are accessed contiguously in GPU memory, thereby enabling fully coalesced memory transactions and significantly reducing the frequency of high-latency global DRAM accesses.
    \item \textbf{Adaptive Fine-Grained Mixed-Precision Strategy:} Leveraging the prior physical knowledge of heat transfer within 3D ICs, we propose an adaptive and fine-grained mixed-precision strategy, which adjusts arithmetic precision across different convergence stages and functional components based on their precision sensitivity. This strategy effectively avoids resource contention and mitigates the scarcity of double-precision units on modern consumer-grade GPUs, thereby enhancing the hardware utilization and computational throughput.
    \item \textbf{High-Parallelism Preconditioning with Synergy:} We adjust the conventional preconditioning strategy and turn to high-parallelism polynomial preconditioning to resolve the observed parallelism-convergence contradiction on GPU architectures. This preconditioning framework also has significant synergistic effects for the proposed Quad-Diag, SpMV kernels, and mixed-precision strategy, thereby maximizing the overall performance.
    \item \textbf{Performance:} Experimental results demonstrate that CUTh-Solver achieves significant speedup over the latest GPU-accelerated COMSOL Multiphysics 6.4, and a wide range of NVIDIA's native general-purpose sparse libraries (AmgX, cuSPARSE, cuDSS).
\end{itemize}

The remainder is organized as follows:
\Cref{sec:2} introduces related works.
\Cref{sec:3} gives basic preliminaries.
\Cref{sec:4} introduces the methodology.
\Cref{sec:5} presents experiments.
\Cref{sec:6} draws the conclusion and future work.

%% file: figs/intro_story.tex
\begin{figure}
    \centering
    \includegraphics[width=1.0\linewidth]{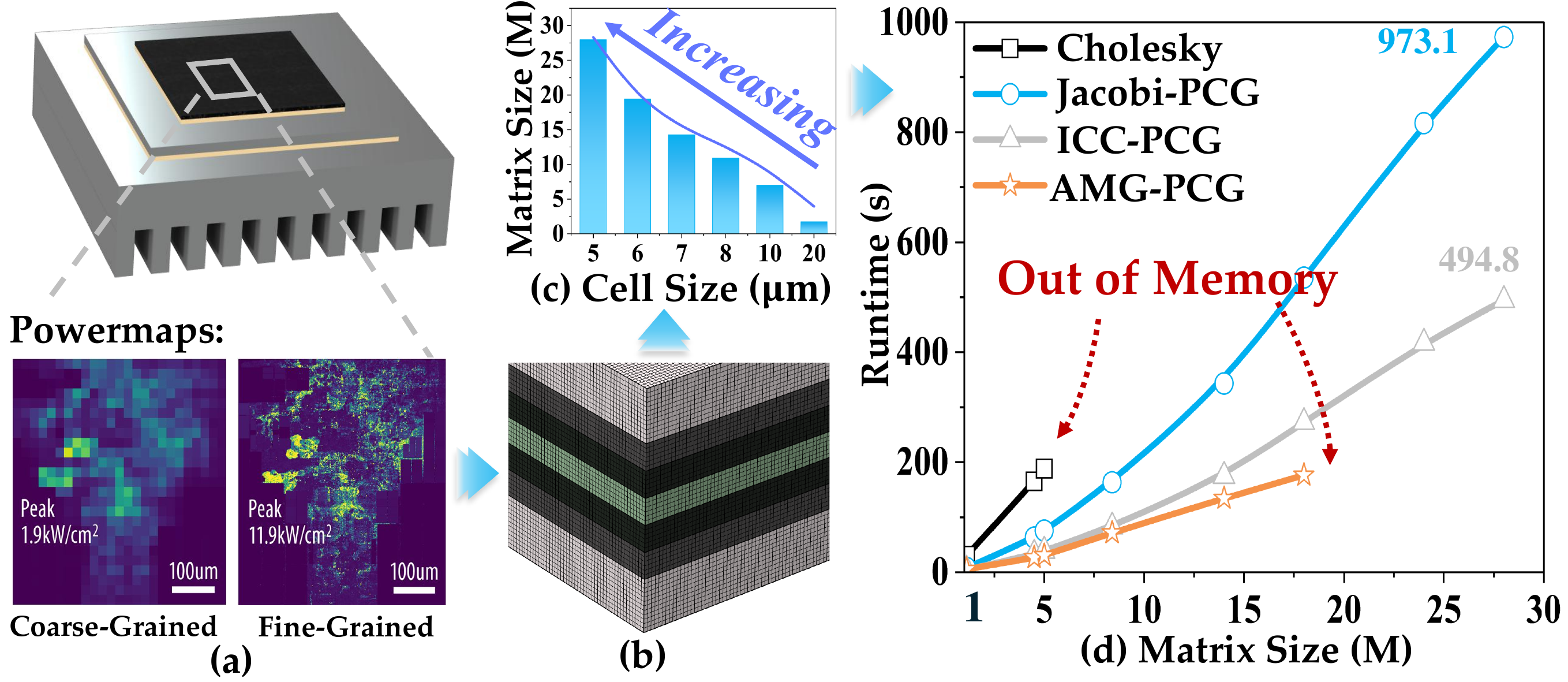}
    \caption{(a) Fine-grained (1-\textmu m) power maps reveal up to $6.3\times$ higher peak power density than coarse-grained (20-\textmu m) ones~\cite{multiscale_iedm}. (b) High-resolution simulation cells are needed to capture fine-grained design features. (c) Matrix size growth with increasingly finer cell size. (d) Efficiency of CPU-based sparse solvers on large-scale systems $\bm{Ax=b}$.}
    \label{intro_story}
\end{figure}

%% file: chapters/2_relatedworks.tex
\section{Related Works}
\label{sec:2}
This section reviews previous work related to sparse linear solvers of thermal simulators on CPUs and GPUs.
\subsection{CPU-Based Sparse Solvers in Thermal Simulators}
Most existing thermal simulation frameworks rely on CPU-based sparse solvers. HotSpot 7.0~\cite{hotspot_7.0}, 3D-ICE 4.0~\cite{3dice4.0}, and MFIT~\cite{MFIT} employ direct solvers through the SuperLU library~\cite{SuperLU-website}. PACT~\cite{PACT} offers multiple solver options through interfaces to KLU~\cite{klu}, AztecOO~\cite{aztecoo}, and Belos~\cite{belos}. ThermalScope~\cite{thermalscope} uses stationary iterative methods such as Gauss-Seidel (GS), while MTA~\cite{MTA_tcad} adopts algebraic multigrid (AMG)-PCG via HYPRE library~\cite{hypre_paper}. The framework in~\cite{cpu_amg, aspdac_amg} similarly employs aggregation-based AMG-PCG. Additionally, the works in~\cite{multigrid_1, multigrid_2} adopt Geometric Multigrid (GMG) as the sparse solver using domain-specific knowledge.

Although these CPU-based solvers are effective at small-to-medium scales, they face significant scalability challenges for high-resolution simulations. Achieving sign-off-grade accuracy within practical turnaround times calls for a shift toward modern GPU architectures to harness its compute power.

\subsection{GPU-Accelerated Sparse Solvers of Thermal Simulators}
 Some research has increasingly turned to GPU acceleration. For example, the work in~\cite{multigrid_gpu} introduced a GPU-accelerated multigrid framework using the ELLPACK storage format, achieving substantial speedups over direct solvers and various Krylov subspace methods. ARTSim 2.0~\cite{boris_tvlsi} accelerates LU factorization using a multifront method to achieve task-level parallelism in GPU architectures.

Despite these valuable contributions, there are opportunities for further optimization. The ELLPACK-based solver in~\cite{multigrid_gpu} achieves coalesced access to matrix elements, but access to the operand vector during SpMV remains irregular. Additionally, its multigrid V-cycle and tridiagonal sub-solver introduce fine-grained data dependencies that limit GPU parallelism, and the proposed smoother is specifically designed for microchannel cooling rather than purely solid-state heat transfer. ARTSim 2.0~\cite{boris_tvlsi} adopts a hybrid CPU-GPU architecture, offloading only a portion of workload to GPUs while retaining symbolic analysis and triangular solves on CPUs, which may limit its scalability for high-resolution 3D IC thermal analysis. Meanwhile, ARTSim 2.0 does not take advantage of domain-specific knowledge for optimization.

%% file: chapters/3_preliminaries.tex
\section{Preliminaries}
\label{sec:3}
This section establishes the necessary background. We first introduce the governing partial differential equations (PDEs) for heat transfer. The corresponding numerical discretization schemes are then detailed in \Cref{pre: numerical_discretization}, followed by an overview of the PCG algorithm in \Cref{pre: PCG_algorithm}.

\subsection{Governing Heat PDEs}
\label{pre: PDEs}
Heat transfer within solid 3D ICs is governed by classic Fourier's Law as the following PDE:
\begin{equation}
    c(x)\frac{\partial T(x,t)}{\partial t}-\nabla \cdot(\kappa(x)\nabla T(x,t)) = p(x,t)\ in\ \mathbf{\Omega}
    \label{heat_pde}
\end{equation}
where $x$ refers to the 3D spatial coordinates $(x_1, x_2, x_3)$, $\kappa(x)$ and $c(x)$ denote space-dependent thermal conductivity and volumetric heat capacity and thermal conductivity, $p(x,t)$ represents the dynamic power maps, and $\mathbf{\Omega}$ describes the simulation domain of 3D ICs.

The Robin convective boundary condition is applied to the convective surfaces $\mathbf{\Gamma_R}$, expressed as follows:
\begin{equation}
    \kappa(x,y)\frac{\partial{T(x,t)}}{\partial{\bm{n}}}|_{\mathbf{\Gamma_R}} = -h(T(x,t) - T_{amb})
    \label{robin_boundary_condition}
\end{equation}
where $h$ is the heat transfer coefficient ($W/m^2K$), and $T_{amb}$ is the air temperature. $\bm{n}$ is the outward normal vector.

The Neumann boundary condition is applied to the adiabatic surfaces $\mathbf{\Gamma_N}$, formulated as follows:
\begin{equation}
    \kappa(x,y)\frac{\partial{T(x,t)}}{\partial{\bm{n}}}|_{\mathbf{\Gamma_N}} = q_{Neumann} = 0
    \label{neumann_boundary_condition}
\end{equation}
where $q_{Neumann}$ refers to the heat flux flowing through $\mathbf{\Gamma_N}$ in Neumman boundary condition.

\Cref{heat_pde}, \Cref{robin_boundary_condition} and \Cref{neumann_boundary_condition} constitute the whole PDEs of heat transfer within solid-state 3D IC.
\input{figs/pre_cells}
\input{figs/overview.tex}

\subsection{Discretization of Heat PDEs}
\label{pre: numerical_discretization}
The standard 7-point finite difference/volume stencil is widely used for spatial discretization in thermal simulation of heat PDEs \cite{PACT, 3dice4.0, hotspot_7.0}. As shown in \Cref{pre_cells} (b), a generic cell is typically adjacent to six neighboring cells. Therefore, the steady-state equation for cell 0 ($c_0$) can be formulated as:
\begin{equation}
    \sum_{i=1}^{6}{g_{0, i}(T_0-T_i) + hS_0T_0= p_0V_0 + hS_0T_{amb}}
    \label{steady-state}
\end{equation}
where $g_{0,i}$ represent the thermal conductance between $c_0$ and $c_i$, $S_0$ refers to its area of $\Gamma_R$ ($S_0$ = 0 since $c_0$ is not connected to the ambient air), and $V_0$ denotes $c_0$'s volume.

The implicit Euler method is also commonly adopted for temporal discretization due to its numerical stability \cite{3dice4.0, MFIT, ETLA-3D}. After temporal discretization, transient heat equation of $c_0$ can be formulated as follows:
\begin{equation}
    \begin{aligned}
        &C_0V_0\frac{T_{0, n+1}}{\Delta t}+\sum_{i=1}^{6}{g_{0, i}(T_{0, n+1}-T_{i, n+1})}+ hS_0T_{0, n+1} \\
        &= p_{0,n+1}V_0 + hS_0T_{amb} + C_0V_0\frac{T_{0,n}}{\Delta t}
    \end{aligned}
    \label{transient}
\end{equation}
where $C_0$ denotes volumetric heat capacity, $\Delta t$ refers to the time step size in transient thermal simulation, $n$ and $n+1$ represent the current and the next time step, respectively.

By applying \Cref{steady-state} and \Cref{transient} across all simulation cells, we construct a global sparse linear system $\bm{Ax=b}$ for both steady-state and transient thermal simulation. Matrix $\bm{A}$ is an SPD banded matrix characterized by a structured sparsity pattern (typically comprising seven parallel non-zero diagonals), as depicted in \Cref{pre_cells} (c), which offers opportunities for domain-specific optimization.

\subsection{Preconditioned Conjugate Gradient (PCG)}
\label{pre: PCG_algorithm}
As described in \Cref{algo:pcg_algorithm}, the PCG method is well-suited to solve $\bm{Ax=b}$ when $\bm{A}$ is SPD. $\bm{M^{-1}}$ is the preconditioner, which is supposed to be an approximate inverse of $\bm{A}$ and easy to be solved. In each CG iteration, the computational overhead is typically dominated by preconditioning and SpMV operation \cite{jinzhou_2024_dac, azul}. 

The implicit preconditioning phase (lines 2 and 10) involves solving an inner linear system $\bm{Mz_{k+1}=r_{k+1}}$ to cluster the eigenvalues to accelerate the convergence. The efficacy of PCG depends on the selection of a preconditioner $\bm{M^{-1}}$ that closely approximates $\bm{A}$ while ensuring that the action of $\bm{M^{-1}}$ is computationally inexpensive to obtain.
\input{algorithms/algorithm_1}

%% file: figs/pre_cells.tex
\begin{figure}
    \centering
    \includegraphics[width=1.0\linewidth]{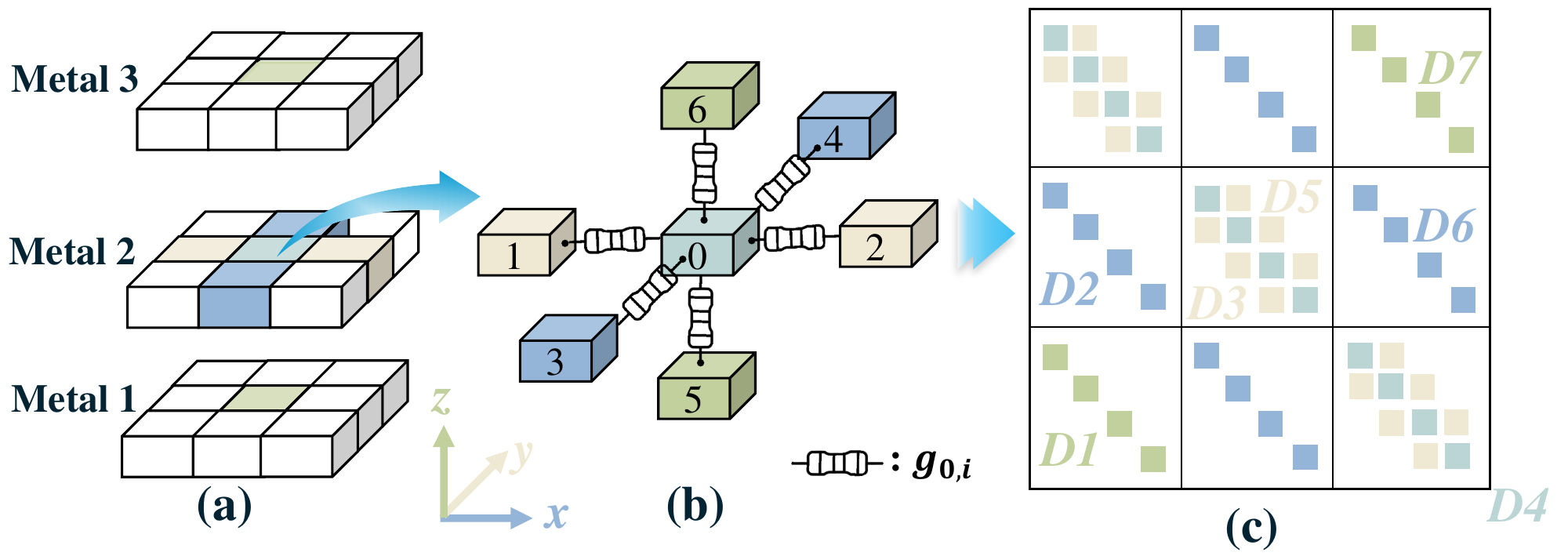}
    \caption{(a) Structured mesh of a 3D IC. (b) Standard 7-point finite difference/volume stencil: each cell interacts with itself and six neighbors. (c) SPD banded matrix with 7 diagonals.}
    \label{pre_cells}
\end{figure}

%% file: figs/overview.tex
\begin{figure*}[t!]
    \centering
    \includegraphics[width=1.0\linewidth]{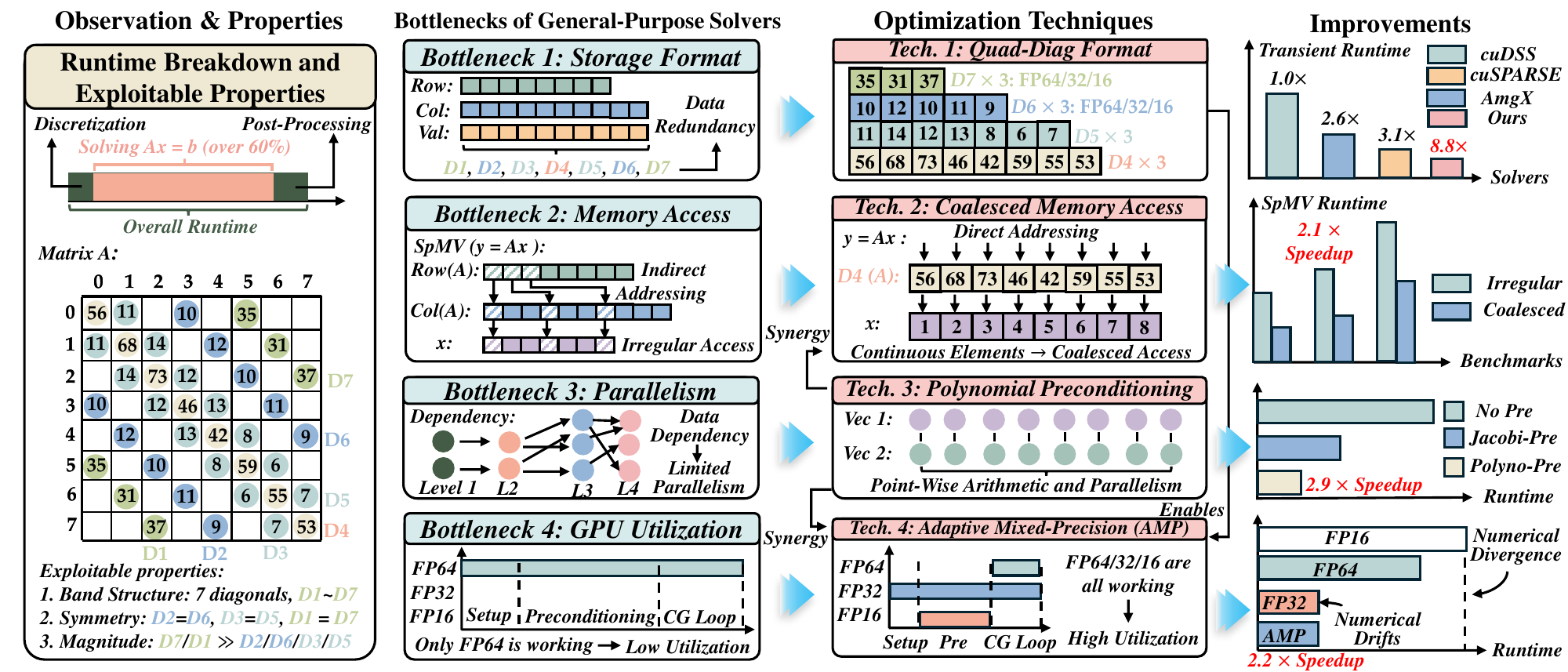}
    \caption{Overview of CUTh-Solver. Exploitable properties of high-resolution 3D IC thermal simulation are identified and exploited. General-purpose solvers that overlook these properties face several bottlenecks. Four optimization techniques are proposed to leverage these opportunities, achieving considerable improvements over general-purpose solvers.}
    \label{fig:overview}
\end{figure*}

%% file: algorithms/algorithm_1.tex
\begin{algorithm2e}
    \SetAlgoLined
    \caption{Preconditioned Conjugate Gradient}
    \KwIn{$\bm{A}, \bm{b}, \bm{M}, \bm{x}_0, \epsilon$}
    \KwOut{$\bm{x}_k$}
    $\bm{r}_0 = \bm{b} - \bm{A}\bm{x}_0$ \tcp{Initial Residual}
    $\bm{z}_0 = \bm{M}^{-1}\bm{r}_0$ \tcp{Preconditioning}
    $\bm{p}_0 = \bm{z}_0$ \tcp{Initialization of $\bm{p}$}
    $k = 0$ \tcp{No. Iteration}
    \While{$\|\bm{r}_k\|_2>\epsilon$}{
        $\bm{\mu}_k = \bm{A}\bm{p}_k$\tcp{Reusable SpMV}
        $\alpha_k = \frac{\bm{r}_k^T \bm{z}_k}{\bm{p}_k^T \bm{\mu}_k}$ \tcp{Two Inner Products}
        $\bm{x}_{k+1} = \bm{x}_k + \alpha_k \bm{p}_k$ \tcp{AXPY}
        $\bm{r}_{k+1} = \bm{r}_k - \alpha_k \bm{\mu}_k$ \tcp{AXPY}
        $\bm{z}_{k+1} = \bm{M}^{-1}\bm{r}_{k+1}$ \tcp{Preconditioning}
        $\beta_k = \frac{\bm{r}_{k+1}^T \bm{z}_{k+1}}{\bm{r}_k^T \bm{z}_k}$ \tcp{Inner Product}
        $\bm{p}_{k+1} = \bm{z}_{k+1} + \beta_k \bm{p}_k$ \tcp{AXPY}
        $k = k+1$ \tcp{Next Iteration}
    }
    \label{algo:pcg_algorithm}
\end{algorithm2e}

%% file: chapters/4_methodology.tex
\section{Methodology: CUTh-Solver on GPUs}
\label{sec:4}
This section presents the methodology of CUTh-Solver. \Cref{meth: overview} provides an overview. \Cref{meth: storage} introduces Quad-Diag format and coalesced memory access in SpMV. \Cref{meth: preconditioning} describes a high-parallelism preconditioning strategy with synergistic effects. Finally, \Cref{meth: mixed-precision} details the mixed-precision scheme within the PCG algorithm.
\input{figs/spmv}

\subsection{Overview of CUTh-Solver}
\label{meth: overview}
\Cref{fig:overview} presents an overview of CUTh-Solver. In numerical thermal simulation, solving $\bm{Ax}=\bm{b}$ dominates the overall runtime (more than 60\%) \cite{nvidia_amgx, taco}. However, general-purpose GPU sparse solvers overlook several domain-specific properties of $\bm{A}$, resulting in four bottlenecks: redundant storage, irregular memory access, limited parallelism, and low GPU utilization.

To address these issues, we propose \textbf{CUTh-Solver} for high-resolution steady-state and transient 3D IC thermal simulation, which exploits these opportunities via four co-designed GPU optimizations: (i) the \textit{Quad-Diag format} eliminates data redundancy; (ii) Diagonal-wise SpMV computation pattern achieves coalesced memory access, thus reducing the frequency of high-latency global DRAM accesses; (iii) Polynomial preconditioning delivers point-wise parallelism and superior synergy than common preconditioners; and (iv) an adaptive mixed-precision strategy leverages heterogeneous compute units to alleviate hardware contention and utilization, while avoiding numerical issues observed in fixed-precision solvers. 

Together, CUTh-Solver achieves significant improvement over state-of-the-art (SOTA) tools and sparse solvers, including the GPU-accelerated COMSOL Multiphysics 6.4, NVIDIA cuDSS, cuSPARSE, and AmgX.

\subsection{Quad-Diag Format and Coalesced Memory Access}
\label{meth: storage}
General-purpose storage formats (CSR, ELLPACK, DIA) suffer from low efficiency of SpMV in domain-specific tasks due to: redundant data storage, irregular memory access, and lack of mixed-precision support. We observe that matrices from 3D IC thermal simulation possess three exploitable properties: SPD symmetry, 7-diagonal band structure, and physical magnitude disparity between diagonals. Leveraging these domain-specific properties of heat transfer within 3D ICs, we propose \textbf{Quad-Diag}, a redundancy-free format, together with four specialized SpMV kernels. This domain-specific design achieves $2.1\times$ SpMV speedup over the general-purpose format and facilitates the subsequent preconditioning and mixed-precision strategies.

Quad-Diag removes data redundancy by condensing the conventional DIA format: It exploits the SPD property of matrix $\bm{A}$ to halve storage, and eliminates the padding overhead inherent in conventional DIA. Padding removal is motivated by our observation that high-resolution 3D IC simulation produces matrices with large diagonal bandwidth, where conventional inter-diagonal alignment incurs substantial memory waste. As illustrated in \Cref{spmv}, by exploiting matrix symmetry, Quad-Diag stores only the upper triangular portion (4 unaligned diagonals: $\bm{D_4}$ through $\bm{D_7}$) in triplicate as $\bm{A_1}$, $\bm{A_2}$, and $\bm{A_3}$ at double-, single-, and half-precision, respectively.
\input{algorithms/algorithm_2}

To achieve coalesced memory access in SpMV kernels, we adopt a diagonal-wise algorithm based on Quad-Diag. As shown in \Cref{spmv}, facilitated by Quad-Diag, the diagonal-wise computation pattern transforms finite inner products into seven element-wise multiplications followed by one vector addition without global reductions. The procedure is formalized in \Cref{diagonal_wise_spmv}, where $\odot$ denotes element-wise multiplication and $ofs$ denotes the column offset of each diagonal relative to the main diagonal. This diagonal-wise SpMV computation enables coalesced memory transactions by accessing seven continuous subsequences within $\bm{x}$, rather than scattered elements in conventional row-wise SpMV computation. By minimizing costly global DRAM accesses, the achieved coalesced memory access pattern leads to a substantial reduction in SpMV execution time.

\subsection{High-Parallelism Preconditioning with Synergy}
\label{meth: preconditioning}
\input{figs/cpu_versus_gpu}
GPU preconditioning suffers from a fundamental conflict between convergence quality and parallelism that is absent on CPUs. CPU-favored preconditioners such as incomplete Cholesky (IC) and algebraic multigrid (AMG) accelerate convergence through complex data dependencies (e.g., SpTRSV, V-cycles), but these dependencies severely limit GPU parallelism. As shown in \Cref{fig:cpu_versus_gpu},  IC-PCG on GPUs can even underperform unpreconditioned CG, indicating that naively migrating CPU-favored techniques to GPUs is suboptimal. Guided by this observation, we adopt the Chebyshev polynomial preconditioner, which (i) is composed exclusively of high-parallelism BLAS operations (SpMV and AXPY) without global reductions or recursive dependencies, (ii) requires no explicit storage of $\bm{M}^{-1}$, and (iii) provides far stronger spectral approximation than its most common high-parallelism alternative, Jacobi~\cite{yousef_textbook}. Beyond its inherent parallelism, Chebyshev preconditioning establishes two unique synergies with the rest of CUTh-Solver, delivering an over $2.5\times$ synergistic acceleration (detailed in \Cref{tab:abl_pre}) that no other preconditioner achieves. Specifically, Chebyshev preconditioning amplifies our other optimizations in two 
ways: 

\textbf{1) SpMV Kernel Dominance:} The preconditioning phase is 
decomposed into a series of SpMV operations, increasing the total 
SpMV count to $N_{cg} \times (k+1)$, where $k$ is the Chebyshev polynomial degree, and making SpMV the dominant kernel of PCG. This allows the solver to fully capitalize on Quad-Diag and the optimized SpMV kernels.

\textbf{2) Mixed-Precision Amplification:} Chebyshev preconditioning 
shifts over $90\%$ of total runtime to the preconditioning phase, 
which only requires a robust estimate of $\bm{z}$ rather than an exact 
result and thus exhibits high numerical resilience to reduced 
precisions since the outer CG iteration corrects residual errors introduced by inexact preconditioning. This creates ample headroom for low-precision compute 
units (e.g., FP16) to be exploited by the mixed-precision strategy 
in \Cref{meth: mixed-precision}. As validated by \Cref{tab:abl_pre}, alternative preconditioners gain less than $1.2\times$ from the same optimization suite, confirming the unique synergy of Chebyshev.

The mathematical principle by which the Chebyshev polynomial preconditioning approximates the action of $\bm{A^{-1}}$ can be explained as follows: 

For CG method with $P_k(0)=1$, the error vector of the $k$-th iteration, $\bm{e_k}$, is formulated as:
\begin{equation}
    \bm{e_k} = \bm{x_{exact}}-\bm{x_k}= P_k(\bm{A})\cdot \bm{e_0}
\end{equation}
where $\bm{e_0}$ denotes the error vector of the initial guess $\bm{x_0}$, and $P_k(\bm{A})$ represents the $k$-degree polynomial of matrix $\bm{A}$. Based on the definition of the matrix 2-norm, the following inequality holds \cite{matrix_analysis}:
\begin{equation}
    \|\bm{e_k}\|_2 \le \|P_k(\bm{A})\|_2\cdot\|\bm{e_0}\|_2
    \label{equa:error_inequality}
\end{equation}
According to the spectral mapping theorem, $P_k(\bm{A})$ maintains the SPD property of $\bm{A}$, as expressed by \cite{matrix_analysis}:
\begin{equation}
    P_k(\bm{A}) = \bm{V}\ diag\left( P_k(\lambda_1), P_k(\lambda_2), \dots, P_k(\lambda_n) \right) \bm{V^T}
\end{equation}
where $\lambda_i$ and $\bm{V}$ denote the eigenvalues and the orthogonal eigenvector matrix of $\bm{A}$, respectively. $diag$ represents diagonal matrix. Therefore, \Cref{equa:error_inequality} can be written as follows:
\begin{equation}
    \|\bm{e_k}\|_2 \le \bm{\max_{i}} |P_k(\lambda_i)|\cdot\|\bm{e_0}\|_2
\end{equation}
To maximize the convergence rate, the objective is to identify a polynomial $P_k(\lambda)$ that satisfies:
\begin{equation}
    \bm{\min_{P_k \in \Pi_k, P_k(0)=1}} \bm{\max_{\lambda \in [\alpha, \beta]}} |P_k(\lambda)|
    \label{equa:objective}
\end{equation}
where $[\alpha, \beta]$ represents the spectral range defined by the minimum and maximum eigenvalues of $\bm{A}$. This is a minimax problem. In approximation theory, its solution is provided by the first-kind Chebyshev polynomial defined as \cite{yousef_textbook}:
\begin{subequations}
\begin{align}
    P_k(\lambda) &= \frac{C_k(t)}{C_k(\theta / \delta)}, \quad t = \frac{\theta - \lambda}{\delta} \\
    \theta &= \frac{\beta + \alpha}{2}, \quad \delta = \frac{\beta - \alpha}{2}
\end{align}
\end{subequations}
where $C_k$ is the degree-$k$ Chebyshev polynomial on the interval $[-1, 1]$, satisfying the three-term recurrence \cite{numerical_analysis}:
\begin{equation}
   C_k(x) = 2x C_{k-1}(x) - C_{k-2}(x),\ C_0(x)=1,\ C_1(x)=x
\end{equation}

By applying this recurrence to the error evolution and utilizing the relationships $\bm{e_k = x_{exact} - x_k}$ and $\bm{r_k = b - Ax_k}$, we derive the iterative update for the solution vector:
\begin{subequations}
\label{equa:final_expression_recurrence}
\begin{align}
    &\bm{x}_{k+1} - \bm{x}_k = \rho_k \left[ \rho_{k-1}(\bm{x}_k - \bm{x}_{k-1}) + \frac{2}{\delta} \bm{r}_k \right] \\
    \rho_k &= \frac{1}{2\sigma_1 - \rho_{k-1}}, \quad \sigma_{k+1} = 2\sigma_1 \sigma_k - \sigma_{k-1}
\end{align}
\end{subequations}
with initial conditions $\sigma_1 = \theta/\delta$, $\sigma_0 = 1$, and $\rho_0 = \delta/\theta$.

\Cref{algo:chebyshev} outlines the Chebyshev polynomial preconditioning within PCG (solving $\bm{Mz_{cg}=r_{cg}}$). Unlike IC or MG, the Chebyshev preconditioner does not require explicit storage of $\bm{M^{-1}}$, significantly reducing GPU memory usage.
\input{algorithms/chebyshev_polynomial}

The spectral range required by \Cref{algo:chebyshev} is estimated using the Lanczos algorithm \cite{matrix_computation}, which reduces $\bm{A}$ to a small tridiagonal matrix $\bm{T}$ whose extreme eigenvalues approximate $\lambda_{min}$ and $\lambda_{max}$. In this algorithm, matrix $\bm{A_2}$ is used for $N$-step Lanczos iterations to construct $\bm{T}$ on GPUs, and the solution of the spectrum of $\bm{T}$ is executed on CPUs.

\subsection{Adaptive Mixed-Precision Strategy in PCG}
\label{meth: mixed-precision}
Modern consumer-grade GPUs increasingly prioritize low-precision throughput over FP64 capability, with FP64-to-FP32 ratios as low as 1:32. Nevertheless, scientific computing conventionally defaults to FP64 for numerical integrity 
\cite{mixed_precision_taco}, forcing all arithmetic to queue for scarce FP64 units while abundant FP32/FP16 units sit idle. However, naively switching to low precision is unsafe: rounding errors can corrupt the Krylov subspace and cause numerical drift, residual stagnation, or even divergence, as empirically confirmed by the fixed-FP16 and fixed-FP32 columns in \Cref{tab:mixed_precision}.

To fully exploit low-precision throughput without sacrificing fidelity, we propose an adaptive mixed-precision PCG that allocates precision along \emph{two 
orthogonal dimensions}, governed by a safety mechanism:
\begin{itemize}
    \item \textbf{Stage-wise (temporal):} the solver is bifurcated into an \emph{early stage} that aggressively uses low precision for fast residual reduction, and a \emph{late stage} that reverts to FP64 for accurate convergence.
    \item \textbf{Component-wise (spatial):} within each stage, precision is assigned granularly to distinct PCG components (preconditioner, SpMV, vector updates, inner products) based on their numerical sensitivity.
    \item \textbf{Safety mechanism:} a tripartite switching criterion triggers the early-to-late transition, followed by a \emph{hot restart} that purges accumulated low-precision errors while preserving the established Krylov subspace.
\end{itemize}
This design delivers a $2.2\times$ speedup over fixed-FP64 with $\|\bm{e}\|_\infty\!\le\!3.1\times 10^{-4}$ (shown in \Cref{tab:mixed_precision}), whereas fixed-FP32 incurs numerical drift and fixed-FP16 diverges. 

The remainder of this subsection details: 1) SpMV kernels supporting diverse floating-point precisions, and 2) the precision management strategy of the PCG solver throughout the entire convergence process, from the early to the late stage.

\textbf{SpMV Kernels Supporting Diverse Precisions:}

\textbf{1) Physics-Aware FP32/16 SpMV with Arithmetic Decoupling (AD):} This kernel applies heterogeneous precision across diagonals based on their physical magnitudes. It uses the matrix $\bm{A_3}$, whose diagonals are stored in FP16, reducing the data movement overhead to $50\%$ of FP32 and $25\%$ of FP64, thereby mitigating the memory wall. After loading, $\bm{D_4}$ and $\bm{D_7}$ are promoted from FP16 to FP32 to preserve arithmetic accuracy, while $\bm{D_5}$ and $\bm{D_6}$ remain in FP16 to maximize low-precision throughput and avoid GPU hardware resource contention. The validity of this approach stems from the physical properties of heat transfer in 3D ICs: $\bm{D_4}$ and $\bm{D_7}$ represent out-of-plane thermal conductance, which exhibits large magnitudes due to ultra-short vertical heat paths within thin stacked layers \cite{ETLA-3D}, and thus dominates the SpMV results. In contrast, $\bm{D_5}$ and $\bm{D_6}$ represent in-plane conductance, which has small magnitudes due to the high aspect ratios of stacked thin layers \cite{ETLA-3D}, which contributes less significantly to the results. This physics-aware precision alignment achieves an optimal balance between performance and accuracy. The kernel algorithm is presented in \Cref{algo:heterogeneous_AD}.
\input{algorithms/algorithm_3}

\textbf{2) FP64/32 SpMV with AD:} This kernel loads $\bm{A_2}$ in FP32 to halve data movement compared to FP64, then promotes values to FP64 for arithmetic. This design alleviates memory bandwidth stress without compromising numerical stability, making it suitable for the high-precision phases of the iterative solver where FP16 cannot meet accuracy requirements.

\textbf{3) Fixed-Precision SpMV:} Standard FP64 and FP32 kernels implemented directly via \Cref{diagonal_wise_spmv}.

Together, these four kernels provide the precision flexibility required by the subsequent mixed-precision management throughout the PCG solver:

\textbf{Component-wise precision in Lanczos algorithm:} The $N$-step Lanczos algorithm is executed on the GPU device to construct the tridiagonal matrix $\bm{T}$ in FP32 precision using the fixed FP32 SpMV kernel. The resulting $\bm{T}$ is then transferred from the GPU device to the CPU host for eigen-decomposition to estimate the spectral range of $\bm{A}$ in FP64 precision.

\textbf{Component-wise precision in preconditioning:} The preconditioning phase approximates $\bm{A}^{-1}$ and inherently tolerates a lower precision than the outer CG loop. However, exclusive FP16 use can trigger instability \cite{block_jacobi_wiley}, particularly given the recursive structure of \Cref{algo:chebyshev}. We therefore adopt FP32 for all vector and scalar operations (lines 1--6, 8--10 of \Cref{algo:chebyshev}), which provides sufficient numerical safeguards at negligible cost. For the dominant SpMV kernel (line 7), we deploy the physics-aware FP32/16 SpMV from \Cref{meth: storage}, leveraging FP16 for data movement while arithmetic decoupling preserves accuracy. The Lanczos spectral estimation runs similarly in FP32 with the fixed FP32 SpMV kernel.

\textbf{Component-wise precision in early CG loops:} During early iterations, the residual decreases rapidly and dominates rounding errors. We employ the FP64/32 SpMV kernel with arithmetic decoupling (line 6 of \Cref{algo:pcg_algorithm}) to halve memory traffic while retaining FP64 arithmetic. The remaining operations stay in FP64, since updates of $\bm{x}$ and $\bm{r}$ steer the convergence and the inner products involve global reductions where limited mantissa bits would amplify 
cancelation errors.

\textbf{Component-wise precision in late CG loops:} As the residual diminishes to magnitudes comparable to low-precision unit roundoff, rounding errors begin to mask the true residual. We thus transition to the fixed-FP64 SpMV kernel to eliminate catastrophic orthogonality loss and restore convergence to the prescribed tolerance.

\textbf{Stage-switching criterion:} The early-to-late transition is 
triggered when \emph{any} of the following holds:
\begin{subequations}
\label{equa:switching}
\begin{align}
    \|\bm{r}_{low}\|_2 \le 1\times10^n,& n = log_{10}(\epsilon)/2
    \label{equa:switch_a}\\
    \|\bm{r}^{k}_{true}\|_2 &> 0.9\,\|\bm{r}^{k-1}_{true}\|_2 
    \label{equa:switch_b}\\
    \|\bm{r}_{true}-\bm{r}_{low}\|_2 &\ge 0.5\,\|\bm{r}_{true}\|_2 
    \label{equa:switch_c}
\end{align}
\end{subequations}
where $\bm{r}_{true}=\bm{b}-\bm{Ax}$ and $\bm{r}_{low}$ is the recursively maintained residual. \Cref{equa:switch_a} detects proximity to convergence. \Cref{equa:switch_b} detects stagnation of the true residual despite a decreasing recursive one. \Cref{equa:switch_c} detects significant drift between them. 
$\bm{r}_{true}$ is periodically re-evaluated in FP64, such as every 20 iterations for steady-state and every 5 for transient simulations, since the previous time step often supplies an initial guess that is already in the late stage.

\textbf{Hot restart:} Because $\bm{r}$ and $\bm{p}$ are updated recursively, low-precision rounding errors accumulate over iterations. Upon switching, a hot restart explicitly recomputes $\bm{r}$ and $\bm{p}$ from their defining equations to purge these errors. The solution vector $\bm{x}$ is carried over directly as a high-quality initial guess, allowing late-stage PCG iterations to continue the solution search without losing the effort invested in the early stage.

%% file: figs/spmv.tex
\begin{figure*}[t!]
    \centering
    \includegraphics[width=1.0\linewidth]{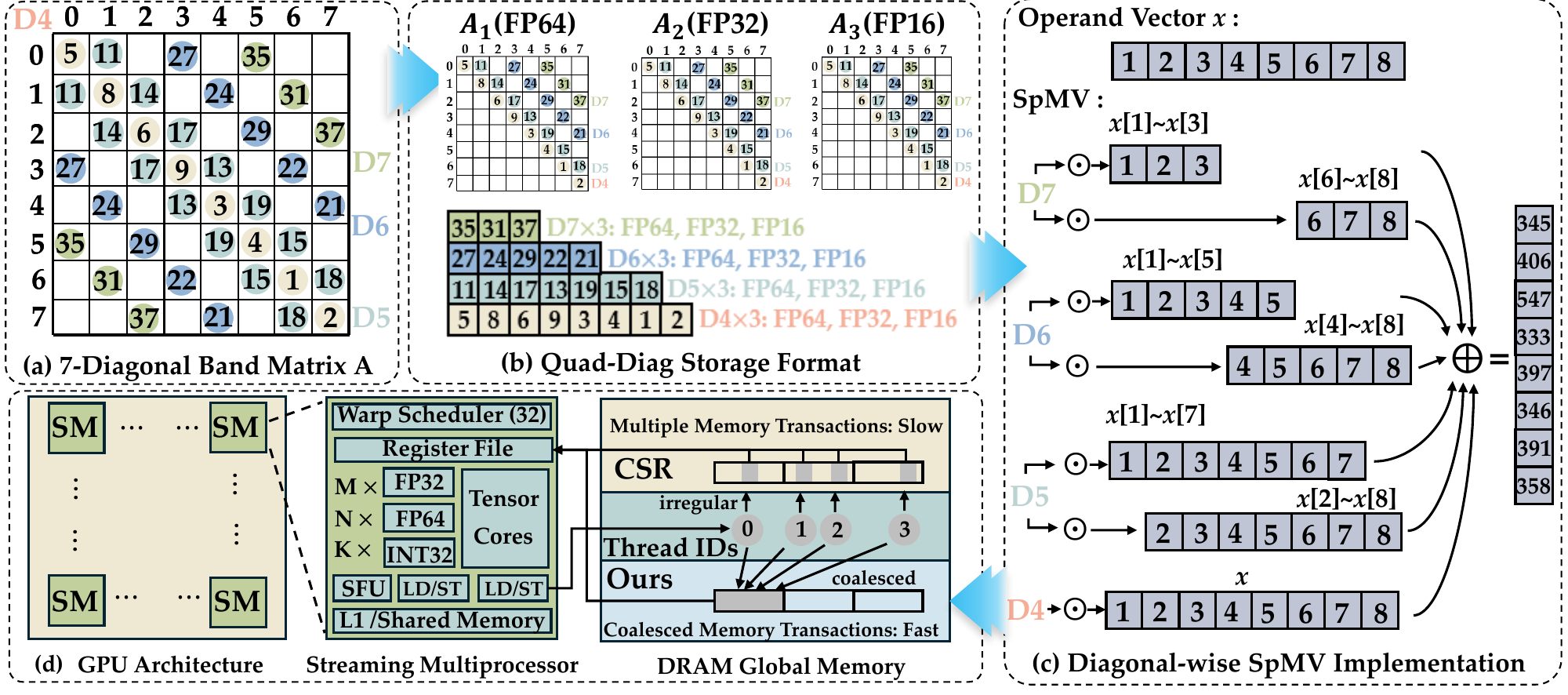}
    \caption{Quad-Diag storage format and diagonal-wise SpMV implementation: (a) a 7-diagonal banded matrix $A$ discretized from 3D IC thermal simulation, (b) Proposed Quad-Diag storage format, (c) Diagonal-wise SpMV implementation, (d)  coalesced memory access to the DRAM global memory enabled by the diagonal-wise SpMV.}
    \label{spmv}
\end{figure*}

%% file: algorithms/algorithm_2.tex
\begin{algorithm2e}[tb!]
    \SetAlgoLined
    \caption{Diagonal-wise SpMV: $\bm{y}=\bm{A}\bm{x}$}
    \label{algo:diagonal_wise_spmv}
    \KwIn{$\bm{D4}$, $\bm{D5}$, $\bm{D6}$, $\bm{D7}$, $N$, $\bm{x}$, $ofs_1$, $ofs_2$}
    \KwOut{$\bm{y}$}
    $\bm{y}=\bm{0}$\tcp{Initialization}
    $\bm{y}\ +=\bm{D4}\odot \bm{x}$\tcp{D4 Component}
    $\bm{y}[1:N-1]\ +=\bm{D5}\odot \bm{x}[2:N]$\tcp{D5}
    $\bm{y}[2:N]\ +=\bm{D5}\odot \bm{x}[1:N-1]$\tcp{D3}
    $\bm{y}[1:N-ofs_1]\ +=\bm{D6}\odot \bm{x}[ofs_1+1:N]$\tcp{D6}
    $\bm{y}[ofs_1:N]\ +=\bm{D6}\odot \bm{x}[1:N-ofs_1]$\tcp{D2}
    $\bm{y}[1:N-ofs_2]\ +=\bm{D7}\odot \bm{x}[ofs_2+1:N]$\tcp{D7}
    $\bm{y}[ofs_2:N]\ +=\bm{D7}\odot \bm{x}[1:N-ofs_2]$\tcp{D1}
    \textbf{Return} $\bm{y}$\\
    \label{diagonal_wise_spmv}
\end{algorithm2e}

%% file: figs/cpu_versus_gpu.tex
\begin{figure}[t!]
    \centering
    \includegraphics[width=1.0\linewidth]{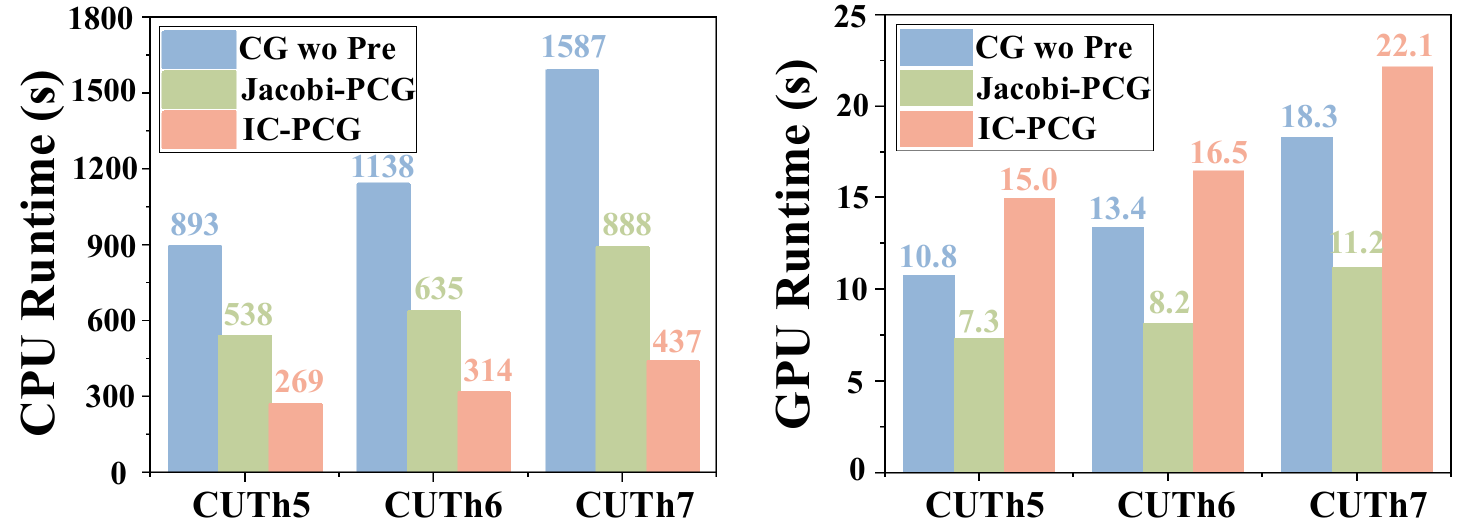}
    \caption{End-to-end runtime of different PCG solvers on CPU and GPU. The performance inversion of IC-PCG highlights the parallelism bottleneck on GPUs. CUThs are benchmarks presented in \Cref{tab: benchmarks}.}
    \label{fig:cpu_versus_gpu}
\end{figure}

%% file: algorithms/chebyshev_polynomial.tex
\begin{algorithm2e}[tb!]
\caption{Chebyshev Polynomial Preconditioning}
\label{algo:chebyshev}
\KwIn{$\bm{A}$, $\bm{r_{cg}}$, $\lambda_{min}$, $\lambda_{max}$, $k$}
\KwOut{$\bm{z_{cg}}$}

\BlankLine
$\bm{r_{chebyshev}}=\bm{r_{cg}}$ \\
$\sigma_1=\theta\ /\ \delta$ \\
$\rho_0=1\ /\ \sigma_1$ \\
$\bm{d_0}=\bm{r_{chebyshev}}\ /\ \theta$ \\
\For{$i = 0$ \KwTo $k-1$}{
    $\bm{z_{cg}} = \bm{z_{cg}} + \bm{d}$ \tcp{AXPY}
    $\bm{Ad} = \bm{A} \cdot \bm{d}$ \tcp{SpMV}
    $\bm{r_{chebyshev}} = \bm{r_{chebyshev}} - \bm{Ad}$ \tcp{AXPY}
    $\rho_{i+1}=1\ /\ (2\sigma - \rho_i)$\tcp{Scalar}
    $\bm{d_{i+1}}=\rho_{i+1}\ \rho_i\ \bm{d_i}+(2\rho_{i+1}\ /\ \delta)\ \bm{r_{chebyshev}}$
}
\Return{$\bm{z_{cg}}$}\\
\end{algorithm2e}

%% file: algorithms/algorithm_3.tex
\begin{algorithm2e}[tb!]
\caption{Physics-Aware FP32/16 SpMV with AD}
\label{algo:heterogeneous_AD}
\SetKwInOut{Input}{Input}
\SetKwInOut{Output}{Output}
\Input{$A_3$ (FP16), $x$ (FP64)}
\Output{$y$ (FP32)}

\BlankLine
$y=0$ \tcp{Initialization}
convert $x \rightarrow x_{FP32}$ \\
convert $x \rightarrow x_{FP16}$ \tcp{one-time convert}

\For{each diagonal $d \in \mathcal{D}$}{
    \If{$d \in \{D_4, D_7\}$ \tcp{Out-of-plane: High}}
    {
        \tcp{Arithmetic Decoupling (AD)}
        Load $val_{FP16}$ of $d$ from $A_3$ \\
        Promote $val_{FP16} \rightarrow val_{FP32}$\\
        $y=y + (val_{FP32} \times x_{FP32})$ \\
    }
    \ElseIf{$d \in \{D_5, D_6\}$\tcp{In-plane: Low}}
    {
        \tcp{Direct FP16 Execution}
        Load $val_{FP16}$ of $d$ from $A_3$\\
        $y_{tmp}=val_{FP16} \times x_{FP16}$\\
        Add($y_{tmp} \rightarrow \mathbf{y}$) \tcp{back to FP32}
    }
}
\Return $\mathbf{y}$\;
\end{algorithm2e}

%% file: chapters/5_experiments.tex
\section{Experiments}
\label{sec:5}
This section presents the experimental setup and extensive comparisons, including three end-to-end comparisons to demonstrate the efficiency of CUTh-Solver and three ablation studies to validate individual contributions.

\input{figs/power_maps}
\input{figs/accuracy_comparison}

\subsection{Experimental Setup}
\textbf{3D IC Configuration:} We evaluate CUTh-Solver on a 10-layer Memory-on-Logic (MoL) F2F 3D IC. From bottom to top, the layers are: logic bulk (10~\textmu m), logic FEOL (2~\textmu m), logic BEOL (10~\textmu m), hybrid bonding (2~\textmu m), memory BEOL (10~\textmu m), memory FEOL (2~\textmu m), memory bulk (200~\textmu m), TIM (100~\textmu m), heat spreader (400~\textmu m), and heat sink (1000~\textmu m). Three couples of power maps (\Cref{fig:powermaps}), widely used as benchmarks~\cite{ETLA-3D, hotspot_7.0, xingwei_dac, tfusion}, are applied to the logic/memory FEOL layers, with $h$ of 1000, 10000, and 8000~$W/m^2K$ on the top surface of the heat sink, respectively.

\textbf{Benchmark Matrices:} The matrices listed in \Cref{tab: benchmarks} are used as benchmarks in \Cref{experiment: 2} through \Cref{experiment: 6}. They are real-world thermal matrices discretized by 3D-ICE ~\cite{3dice4.0}, covering both steady-state and transient scenarios. Transient matrices are formed by adding a diagonal heat-capacity term to the steady-state counterparts via Backward Euler, so both share the same dimension ($N$) and number of non-zeros ($NNZ$). The stopping criterion for all iterative solvers is identical to $\|\bm{r_k}\|_2 \leq \epsilon = 1.0e-6$. The Chebyshev polynomial degree $k$ of CUTh-Solver is set to 30.
\input{tables/table_benchmarks.tex}

\textbf{Platform and Baselines:} All experiments are conducted on an Intel Xeon w3-2423 CPU with 32~GB RAM and a consumer-level NVIDIA GeForce RTX 5090 GPU with 32~GB VRAM. The identical memory capacity ensures fair comparison between CPU and GPU methods. CUTh-Solver is implemented in PyTorch 2.9.0. Baselines in \Cref{experiment: 1} include the latest GPU-accelerated COMSOL Multiphysics~\cite{comsol_gpu_acceleration}, HotSpot 7.0 ~\cite{hotspot_7.0}, and 3D-ICE 4.0 ~\cite{3dice4.0}. After tuning, the number of threads for 3D-ICE 4.0 is set to 4 for the best efficiency. The baselines in \Cref{experiment: 2} and \Cref{experiment: 3} include IC(0)-PCG (PETSc 3.24.3 CPU Version~\cite{petsc_paper}), Cholesky (NVIDIA cuDSS 0.5.0~\cite{cuDSS-website, cuDSS-github}), IC(0)-PCG (NVIDIA cuSPARSE 12.5.4~\cite{cuSPARSE-wersite, cuSPARSE-github}), and AMG-PCG (NVIDIA AmgX 2.5.0~\cite{nvidia_amgx, amgx-website, amgx-github}).

\textbf{Notation and Metrics:} Experiments in \Cref{experiment: 2} to \Cref{experiment: 6} adopt the following notations and metrics. Simulation accuracy is measured by MaxAE, MAE, and $\|\bm{e}\|_{\infty}$, denoting the maximum absolute error (\textcelsius), the mean absolute error (\textcelsius), and infinity norm of the error vector, respectively, where
\begin{equation}
\|\bm{e}\|_{\infty}=\|\bm{x}_{\text{exact}}-\bm{x}_{\text{evaluated}}\|_{\infty}=\max_{i}|e_i|.
\end{equation}
The simulation efficiency is characterized by $N_{\text{ite}}$ and $N_{\text{SpMV}}$, i.e., the iteration count and the SpMV execution count of the iterative solvers, together with their SpMV runtime $T_{\text{SpMV}}$, setup time $T_{\text{setup}}$, solve time $T_{\text{sol}}$, and total runtime $T_{\text{tot}}$ (all in seconds).

\subsection{End-to-End Comparison with Thermal Simulators}
\label{experiment: 1}
This subsection presents end-to-end steady-state comparisons between CUTh-Solver and other thermal simulators. The grid resolutions range from $128^2\times10$ to $512^2\times10$ for each couple of power maps. The comparison results are shown in \Cref{tab: e2e_ts} and \Cref{thermal_maps}.
\input{tables/table_e2e_ts}

COMSOL Multiphysics 6.4 results serve as ground truth. With GPU acceleration via NVIDIA cuDSS~\cite{cuDSS-website}, COMSOL consistently outperforms HotSpot 7.0 and 3D-ICE 4.0, both of which rely on CPU-based SuperLU~\cite{SuperLU-website} and encounter Out-of-Memory (OOM) errors at high resolutions. COMSOL further benefits from a hybrid RAM-VRAM memory architecture, whereas SuperLU is limited to host-side memory. 3D-ICE 4.0 exhibits superior efficiency than HotSpot 7.0 since the latest update integrates OpenMP and SuperLU MT for CPU multi-thread execution \cite{3dice4.0}.

CUTh-Solver achieves speedups of up to $10.28\times$, $22.10\times$, and $51.04\times$ over COMSOL at resolutions of $128^2\times10$, $256^2\times10$, and $512^2\times10$, respectively. The average speedup over COMSOL is $25.8\times$. MaxAE and MAE of CUTh-Solver are maintained below 0.20 and 0.06 \textcelsius{}, respectively. No OOM errors were encountered across all resolutions, confirming its efficiency, scalability, and numerical fidelity.

\subsection{End-to-End Comparison with Sparse Solvers: Steady-State}
\label{experiment: 2}

Due to memory constraints observed in the previous comparison, this subsection focuses on end-to-end comparisons using public sparse linear solvers for steady-state problems. For benchmarks CUTh$1\sim4$, NVIDIA cuDSS results serve as ground truth. For the larger cases CUTh$5\sim7$, PETSc IC(0)-PCG results are regarded as ground truth. Comparison results are presented in \Cref{tab: e2e_ss}.
\input{tables/table_e2e_ss_ss}

The Cholesky direct solver (NVIDIA cuDSS) exhibits the poorest scalability, encountering OOM earlier than iterative methods. Its one-time setup phase (symbolic analysis and numerical factorization) dominates the total cost, making it even slower than CPU-based IC(0)-PCG for single-RHS problems. However, since the solve phase itself is inexpensive, direct solvers remain advantageous for multiple-RHS scenarios where setup cost can be amortized.

The GPU-accelerated IC(0)-PCG (cuSPARSE) achieves $11.4\sim20.2\times$ speedup ($16.8\times$ in average) over its CPU counterpart (PETSc), yet remains the least efficient among GPU iterative methods. This is primarily due to the sequential nature of incomplete Cholesky factorization and triangular substitutions, which limit parallelism on SIMT architectures~\cite{yousef_gpu}.

We also include aggregation-based AMG-PCG from NVIDIA AmgX for its relatively low setup cost and effectiveness in thermal simulation. AmgX achieves $20.6\sim40.2\times$ speedup ($32.5\times$ in average) over the CPU baseline and the lowest iteration count, reflecting AMG's superior convergence. However, the per-iteration cost of hierarchical relaxations and prolongations partially offsets its convergence advantage.

CUTh-Solver achieves $52.1\sim70.9\times$ speedup ($59.5\times$ on average) over the CPU baseline with error $\|\bm{e}\|_{\infty}\leq3.3\times10^{-4}$. Its setup cost (Lanczos algorithm) remains at the millisecond level. The high-degree Chebyshev polynomial preconditioner yields fewer iterations than IC(0)-PCG while offering superior parallelism. Although AMG-PCG provides stronger mathematical conditioning, the lower per-iteration cost of CUTh-Solver significantly compensates, resulting in shorter overall wall-clock time across all benchmarks.

\subsection{End-to-End Comparison with Sparse Solvers: Transient}
\label{experiment: 3}

This subsection presents the end-to-end transient comparison, where $\bm{Ax=b}$ is solved across 50 time steps using backward Euler. The setup phase is performed once and reused for all steps. The results are summarized in \Cref{tab: e2e_tr}.
\input{tables/table_e2e_ss_tr}

Unlike the steady-state case, Cholesky (NVIDIA cuDSS) achieves $5.9\sim10.7\times$ speedup ($8.4\times$ in average) over the CPU baseline before OOM, as the factorization cost is amortized across multiple time steps, reducing each subsequent solve to efficient triangular substitutions.

Another notable difference is that IC(0)-PCG (cuSPARSE) slightly outperforms AMG-PCG (AmgX), achieving $16.1\sim34.3\times$ ($25.9\times$ on average) versus $14.9\sim27.3\times$ ($21.8\times$ on average) speedup. As the system approaches thermal equilibrium, the solution between consecutive steps changes minimally, providing near-optimal initial guesses. In this regime, IC(0)-PCG converges within a few low-cost iterations, whereas AMG's higher per-iteration cost offsets its stronger convergence properties.

CUTh-Solver achieves $53.8\sim103.4\times$ speedup ($73.6\times$ in average) over the CPU baseline with $\|\bm{e}\|_{\infty}\leq2.7\times10^{-4}$, delivering up to $3\times$ improvement against both cuSPARSE and AmgX. This advantage stems from a favorable balance between preconditioning efficacy and per-iteration cost: In the early stages, its preconditioning ensures faster convergence than IC(0)-PCG; in later stages, its lower per-iteration cost provides superior throughput compared to AMG-PCG.

\subsection{Ablation Study I: Quad-Diag and Coalesced SpMV}
\label{experiment: 4}
This subsection isolates the contributions of the Quad-Diag format and coalesced SpMV. We benchmark CUTh-Solver against a baseline PCG solver using standard CSR format with row-wise (irregular) SpMV. The adaptive mixed-precision strategy is disabled for both solvers to directly measure the gains from data-layout and memory-access optimizations.

\Cref{tab:abl_spmv} presents the comparison results. Under identical $N_{SpMV}$, Quad-Diag with coalesced kernels delivers a consistent $2\sim2.2\times$ speedup over the CSR baseline in SpMV time across all benchmarks. This kernel-level gain translates into a $1.4\sim1.7\times$ speedup for the overall solve phase, as SpMV operations account for $70\sim85\%$ of total solve time in the Chebyshev-preconditioned PCG solver.
\input{tables/table_ablation_spmv}

The performance gain originates from transforming the irregular memory access patterns of CSR into coalesced accesses, drastically reducing global memory transactions and mitigating the overhead of high-latency DRAM operations. This optimization is particularly effective because SpMV is fundamentally memory-bound~\cite{2020_wiley}: its low arithmetic intensity means execution time is dominated by data movement rather than computation. Additionally, the direct addressing of Quad-Diag eliminates auxiliary index arrays, further improving instruction throughput.

\subsection{Ablation Study II: Adaptive Mixed-Precision Strategy}
\label{experiment: 5}
This subsection quantifies the contribution of the adaptive mixed-precision strategy by benchmarking it against fixed single-precision (FP32) and fixed half-precision (FP16) Chebyshev-preconditioned PCG solvers. Solutions from the fixed double-precision (FP64) solver serve as ground truth.

\input{tables/table_ablation_mixed}
\input{tables/table_ablation_preconditioning}
As shown in \Cref{tab:mixed_precision}, the fixed FP16 solver diverges across all benchmarks. The fixed FP32 solver maintains the same iteration count and delivers a $2.0\sim2.7\times$ speedup over FP64, but introduces a considerable numerical drift of $\|\bm{e}\|_{\infty}=0.64$, corresponding to a $0.64$ \textcelsius\ deviation in the temperature field. In contrast, our mixed-precision strategy achieves solve time comparable to fixed FP32 while maintaining negligible errors ($\|\bm{e}\|_{\infty}\leq3.1\times10^{-4}$). These results demonstrate that reckless use of low-precision arithmetic can cause significant numerical drift or total divergence in scientific computing.

The numerical drift observed in the fixed FP32 solver originates from a low-precision representation of numerically sensitive quantities within the CG loop, particularly the residual vector $\bm{r}$. Rounding errors in $\bm{r}$ accumulate over iterations and corrupt the search directions in the Krylov subspace. Our adaptive strategy mitigates this by selectively employing FP64 for these critical components, preserving orthogonalization integrity. The incremental FP64 overhead is offset by the physics-aware heterogeneous FP16/32 SpMV kernel used in the preconditioning phase, yielding a solve time comparable to fixed FP32 with numerical accuracy comparable to FP64.

\subsection{Ablation Study III: Polynomial Preconditioning}
\label{experiment: 6}
This subsection presents an ablation study to evaluate the synergistic effects between the Chebyshev polynomial preconditioner and our proposed optimization framework (the Quad-Diag format, coalesced SpMV kernels, adaptive mixed-precision strategy). We establish two baseline configurations by replacing the Chebyshev preconditioner with either a Jacobi preconditioner or an identity matrix (no preconditioning). These two preconditioners were selected because they integrate seamlessly with our optimization framework. The optimization framework can be turned on/off to offer a clear demonstration of each solver's synergy.

The results are summarized in \Cref{tab:abl_pre}. In the absence of the optimization framework, the Jacobi-preconditioned CG achieves a $1.43\sim1.75\times$ speedup ($1.60\times$ in average) over the vanilla CG solver, whereas the Chebyshev preconditioner yields a modest $1.06\sim1.21\times$ speedup ($1.12\times$ in average), notably underperforming the Jacobi baseline. However, this performance hierarchy is completely inverted once the optimization framework is enabled. Under the optimized configuration, the Chebyshev preconditioner achieves a substantial $2.20\sim3.25\times$ speedup ($2.86\times$ in average) over the vanilla CG solver, while the Jacobi-PCG gain is limited to $1.50\sim1.91\times$ ($1.74\times$ on average). Specifically, activating the optimization suite delivers an improvement in performance over $2.5\times$ for the Chebyshev preconditioner, underscoring a significant synergistic effect. In contrast, the additional gains for the vanilla CG and Jacobi-PCG remain below $1.24\times$ and $1.12\times$, respectively. These results demonstrate that the Chebyshev preconditioner, owing to its compute-intensive and SpMV-centric nature, uniquely capitalizes on our proposed optimization suite. In contrast, SpMV and preconditioning are less dominant in the vanilla CG and Jacobi-PCG solvers, so they benefit less from our optimization framework. 


%% file: figs/power_maps.tex
\begin{figure}[t!]
    \centering
    \includegraphics[width=1.0\linewidth]{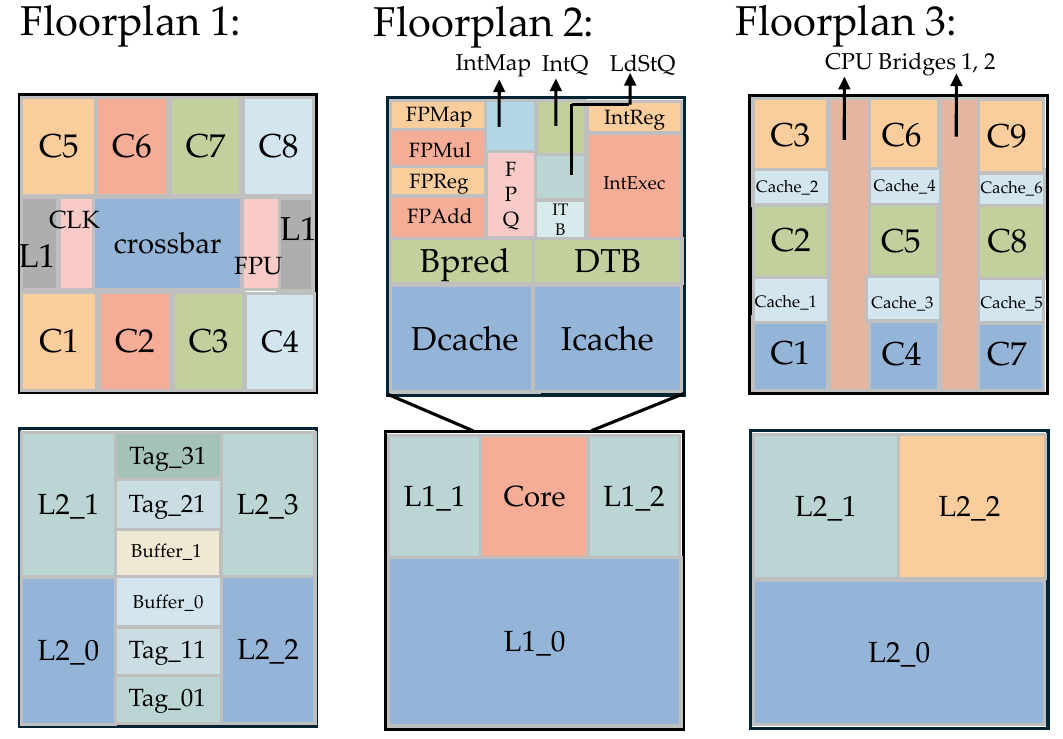}
    \caption{Three MoL floorplans: Logic floorplan 1 is the SPARC processor, logic floorplan 2 is the EV6 processor with the core magnified, and logic floorplan 3 is a simplified multi-core processor. Floorplans 2 and 3 share the same 3-cache memory (right bottom).}
    \label{fig:powermaps}
\end{figure}

%% file: figs/accuracy_comparison.tex
\begin{figure*}[t!]
    \centering
    \includegraphics[width=1.0\linewidth]{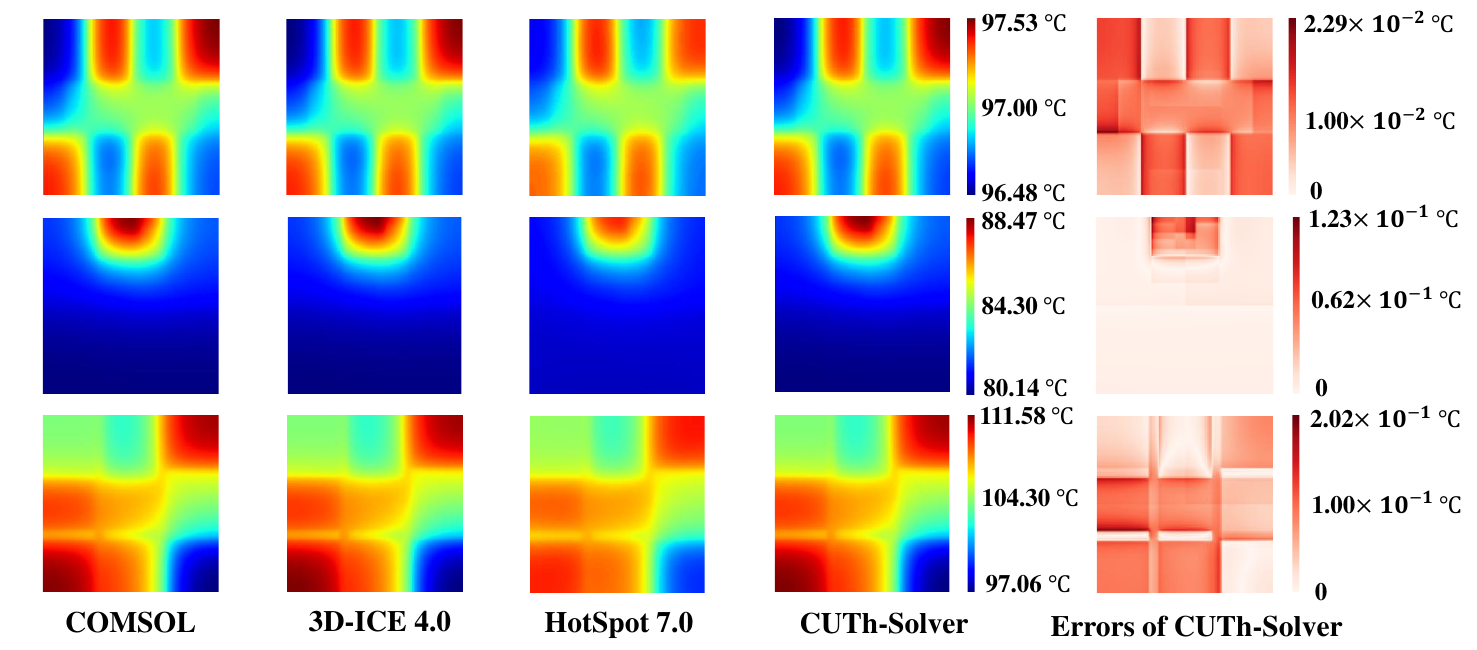}
    \caption{Thermal maps of the 3D IC top tier by COMSOL Multiphysics \cite{comsol_gpu_acceleration}, 3D-ICE 4.0\cite{3dice4.0}, HotSpot 7.0\cite{hotspot_7.0}, CUTh-Solver, and the error map of CUTh-Solver. Three cases  from the top to the bottom are SPARC, EV6, and 9-Core, respectively.}
    \label{thermal_maps}
\end{figure*}

%% file: tables/table_benchmarks.tex
\begin{table}[t!]
    \centering
    \caption{Real-world steady-state and transient matrices discretized from 3D-ICE \cite{3dice4.0}. They are used as matrix benchmarks from \Cref{experiment: 2} to \Cref{experiment: 6}.}
    \label{tab: benchmarks}
    \resizebox{\linewidth}{!}
    {
    \begin{tabular}{ccc}
        \toprule
        Matrix Benchmarks&Dimension ($N$)&Num of Non-Zeros ($NNZ$)\\
        \midrule
        CU\_Therm\_1 (CUTh1)&$5.1\times10^6$&$3.4\times10^7$ \\
        CU\_Therm\_2 (CUTh2)&$8.4\times10^6$&$5.6\times10^7$ \\
        CU\_Therm\_3 (CUTh3)&$1.4\times10^7$&$9.4\times10^7$ \\
        CU\_Therm\_4 (CUTh4)&$1.8\times10^7$&$1.2\times10^8$ \\
        CU\_Therm\_5 (CUTh5)&$2.4\times10^7$&$1.6\times10^8$ \\
        CU\_Therm\_6 (CUTh6)&$2.8\times10^7$&$1.9\times10^8$ \\
        CU\_Therm\_7 (CUTh7)&$3.5\times10^7$&$2.3\times10^8$ \\
        \bottomrule
    \end{tabular}
    }
\end{table}

%% file: tables/table_e2e_ts.tex
\begin{table*}[t!]
    \centering
    \caption{The steady-state performance of GPU-accelerated COMSOL Multiphysics 6.4 \cite{comsol_gpu_acceleration}, HotSpot 7.0 \cite{hotspot_7.0}, 3D-ICE 4.0 \cite{3dice4.0}, and CUTh-Solver. Benchmark is a 10-layer hybrid bonding F2F 3D ICs with 3 different couples of power maps as illustrated in \Cref{fig:powermaps}. The simulation results of COMSOL Multiphysics are considered as the ground truth. OOM refers to out of memory. The memory capacity is 32GB for both RAM and VRAM.}
    \label{tab: e2e_ts}
    \resizebox{\textwidth}{!}
    {
    \begin{tabular}{m{1.5cm}c||c|ccc|ccc||ccc}
    
        \hline
        \multirow{2}{*}{Floorplans}&\multirow{2}{*}{Grid Resol.}&\multicolumn{1}{c|}{COMSOL\cite{comsol_gpu_acceleration}}&\multicolumn{3}{c|}{HotSpot 7.0 \cite{hotspot_7.0}}&\multicolumn{3}{c||}{3D-ICE 4.0 \cite{3dice4.0}}&\multicolumn{3}{c}{CUTh-Solver} \\
    
        &&Runtime(s)& MaxAE & MAE & Runtime(s) & MaxAE & MAE & Runtime(s) & MaxAE & MAE & Runtime(s) \\
        \hline
        \hline
    
        \multirow{3}{*}{\centering SPARC}&$128^2\times10$&11.0 &0.15&0.03&163.5&0.02&0.01&12.4&\textbf{0.02}&\textbf{0.01}&\textbf{1.2} \\
        
        &$256^2\times10$&46.0&-&-&OOM&0.02&0.01&94.7&\textbf{0.02}&\textbf{0.01}&\textbf{2.3}\\

        &$512^2\times10$&176.0&-&-&OOM&-&-&OOM&\textbf{0.02}&\textbf{0.01}&\textbf{3.7}\\
        \hline

        \multirow{3}{*}{\centering EV6}&$128^2\times10$&11.0 &0.96&0.33&162.3&0.12&0.01&19.5&\textbf{0.12}&\textbf{0.01}&\textbf{1.1} \\

        &$256^2\times10$&40.0&-&-&OOM&0.13&0.03&100.6&\textbf{0.13}&\textbf{0.03}&\textbf{1.8}\\

        &$512^2\times10$&179.0&-&-&OOM&-&-&OOM&\textbf{0.13}&\textbf{0.03}&\textbf{3.3}\\
        \hline
    
        \multirow{3}{*}{9-Core}&$128^2\times10$&11.0&$2.38$&0.56&162.1&0.20&0.06&17.9 &\textbf{0.20}&\textbf{0.06}&\textbf{1.2} \\

        &$256^2\times10$&43.0&-&-&OOM&\textbf{0.15}&0.03&102.1&0.16&\textbf{0.03}&\textbf{2.47}\\

        &$512^2\times10$&171.0&-&-&OOM&-&-&OOM&\textbf{0.16}&\textbf{0.03}&\textbf{3.4}\\
        \hline
        \hline

        \multicolumn{2}{c||}{Average Speedup (Avg.Spd)}&$1.0\times$&-&-&$0.1\times$&-&-&$0.6\times$&-&-&\bm{$25.8\times$}\\
        \hline
    \end{tabular}
    }
\end{table*}

%% file: tables/table_e2e_ss_ss.tex
\begin{table*}[t!]
    \centering
    \caption{The performance for one-time solving $\bm{Ax=b}$ from steady-state thermal simulation of 3D ICs. Four baselines are: IC(0)-PCG (PETSc 3.24.3 \cite{petsc_paper} on CPU), Cholesky direct solver (NVIDIA cuDSS 0.5.0 \cite{cuDSS-website, cuDSS-github} on GPU), IC(0)-PCG (NVIDIA cuSPARSE 12.5.4 \cite{cuSPARSE-wersite, cuSPARSE-github} on GPU), aggregation-based AMG-PCG (NVIDIA AmgX 2.5.0 \cite{amgx-website, amgx-github, nvidia_amgx} on GPU). BMs refers to the matrix benchmarks presented in \Cref{tab: benchmarks}. Stopping criterion: $\|\bm{r_k=b-Ax_k}\|_2\leq\epsilon=1.0e-6$.}
    \label{tab: e2e_ss}
    \resizebox{\textwidth}{!}
    {
    \begin{tabular}{c||cccc|ccc|cccc|cccc||ccccc}
    
        \hline
        \multirow{2}{*}{BMs}&\multicolumn{4}{c|}{IC(0)-PCG (PETSc) \cite{petsc_paper}}&\multicolumn{3}{c|}{Cholesky (cuDSS)\cite{cuDSS-github}}&\multicolumn{4}{c|}{IC(0)-PCG (cuSPARSE)\cite{cuSPARSE-github}}&\multicolumn{4}{c||}{AMG-PCG (AmgX)\cite{amgx-github}}&\multicolumn{5}{c}{CUTh-Solver}\\
        &$N_{ite}$&$T_{setup}$&$T_{sol}$&$T_{tot}$&$T_{setup}$&$T_{sol}$&$T_{tot}$&$N_{ite}$&$T_{setup}$&$T_{sol}$&$T_{tot}$&$N_{ite}$&$T_{setup}$&$T_{sol}$&$T_{tot}$&$N_{ite}$&$T_{setup}$&$T_{sol}$&$T_{tot}$&$\|\bm{e}\|_{\infty}$\\
        \hline
        \hline
        CUTh1&199&0.6&25.3&25.9&47.3&0.1&47.4&198&0.5&1.8&2.3&\textbf{4}&0.3&1.0&1.3 &44&\textbf{0.2}&\textbf{0.3}&\textbf{0.5}&\textbf{1.5e-4}\\

        CUTh2&250&1.6&55.8&57.5&86.2&0.1&86.3&249&1.0&2.8&3.8&\textbf{5}&0.3&2.0&2.3 &56&\textbf{0.2}&\textbf{0.6}&\textbf{0.8}&\textbf{2.3e-4}\\

        CUTh3&317&1.8&112.2&114.0&215.6&0.5&216.1&316&2.1&4.8&6.9&\textbf{5}&0.4&3.0&3.4 &71&\textbf{0.3}&\textbf{1.5}&\textbf{1.8}&\textbf{2.3e-4}\\

        CUTh4&351&2.2&159.6&161.8&317.9&1.1&373.0&349&3.1&6.5&9.6&\textbf{6}&0.4&4.9&5.3 &76&\textbf{0.4}&\textbf{2.3}&\textbf{2.7}&\textbf{2.3e-4}\\

        CUTh5&432&3.0&265.3&268.3&-&-&OOM&430&5.0&10.2&15.2&\textbf{6}&0.5&6.4&6.9 &91&\textbf{0.5}&\textbf{4.2}&\textbf{4.7}&\textbf{2.9e-4}\\

        CUTh6&433&3.3&312.0&315.3&-&-&OOM&430&5.8&10.3&16.1&\textbf{6}&0.6&7.4&7.8 &96&\textbf{0.5}&\textbf{5.2}&\textbf{5.7}&\textbf{3.3e-4}\\

        CUTh7&482&4.4&433.3&437.7&-&-&OOM&480&8.0&13.6&21.6&\textbf{7}&0.8&10.6&11.5 &107&\textbf{0.6}&\textbf{7.1}&\textbf{7.7}&\textbf{2.5e-4}\\
        \hline
        \hline
    
        Avg.Spd&-&-&-&$1.0\times$&-&-&$0.6\times$&-&-&-&$16.8\times$&-&-&-&$32.5\times$&-&-&-&$\mathbf{59.5\times}$&-\\
        \hline
    \end{tabular}
    }
\end{table*}

%% file: tables/table_e2e_ss_tr.tex
\begin{table*}[t!]
    \centering
    \caption{The performance for 50-time solving $\bm{Ax=b}$ from transient thermal simulation of 3D ICs. Four baselines are: IC(0)-PCG (PETSc 3.24.3 \cite{petsc_paper} on CPU), Cholesky direct solver (NVIDIA cuDSS 0.5.0 \cite{cuDSS-website, cuDSS-github} on GPU), IC(0)-PCG (NVIDIA cuSPARSE 12.5.4 \cite{cuSPARSE-wersite, cuSPARSE-github} on GPU), aggregation-based AMG-PCG (NVIDIA AmgX 2.5.0 \cite{amgx-website, amgx-github, nvidia_amgx} on GPU). Stopping criterion for a single-step solution: $\|\bm{r_k=b-Ax_k}\|_2\leq\epsilon=1.0e-6$.}
    \label{tab: e2e_tr}
    \resizebox{\textwidth}{!}
    {
    \begin{tabular}{c||cccc|ccc|cccc|cccc||ccccc}
    
        \hline
        \multirow{2}{*}{BMs}&\multicolumn{4}{c|}{IC(0)-PCG (PETSc)\cite{petsc_paper}}&\multicolumn{3}{c|}{Cholesky (cuDSS)\cite{cuDSS-github}}&\multicolumn{4}{c|}{IC(0)-PCG (cuSPARSE)\cite{cuSPARSE-github}}&\multicolumn{4}{c||}{AMG-PCG (AmgX)\cite{amgx-github}}&\multicolumn{5}{c}{CUTh-Solver}\\
        &$N_{ite}$&$T_{setup}$&$T_{sol}$&$T_{tot}$&$T_{setup}$&$T_{sol}$&$T_{tot}$&$N_{ite}$&$T_{setup}$&$T_{sol}$&$T_{tot}$&$N_{ite}$&$T_{setup}$&$T_{sol}$&$T_{tot}$&$N_{ite}$&$T_{setup}$&$T_{sol}$&$T_{tot}$&$\|\bm{e}\|_{\infty}$\\
        \hline
        \hline

        CUTh1&3900&0.6&507.6&508.2&48.6&6.8&55.4&3900&0.7&30.8&31.5&\textbf{125}&0.3&33.7&34.0&791&\textbf{0.2}&\textbf{6.3}&\textbf{6.5}&\textbf{2.5e-4}\\

        CUTh2&4716&1.0&1022.5&1023.5&83.9&12.0&95.8&4716&1.2&48.3&49.5&\textbf{134}&0.3&57.5&57.8&948&\textbf{0.2}&\textbf{9.7}&\textbf{9.9}&\textbf{2.5e-4}\\

        CUTh3&5740&2.0&2074.2&2076.2&214.0&47.2&261.2&5740&2.4&83.9&86.9&\textbf{146}&0.3&98.3&98.6&1166&\textbf{0.3}&\textbf{24.3}&\textbf{24.6}&\textbf{2.7e-4}\\

        CUTh4&6337&2.3&2919.0&2921.3&380.7&113.7&494.4&6335&3.3&106.8&110.1&\textbf{150}&0.3&130.3&130.6&1243&\textbf{0.4}&\textbf{39.4}&\textbf{39.8}&\textbf{2.6e-4}\\

        CUTh5&7542&2.8&4572.4&4574.2&-&-&OOM&7558&5.3&161.8&167.1&\textbf{157}&0.4&188.3&188.7&1830&\textbf{0.4}&\textbf{84.3}&\textbf{84.7}&\textbf{0.5e-4}\\

        CUTh6&7544&3.1&5467.0&5470.1&-&-&OOM&7550&6.2&169.7&175.9&\textbf{156}&0.5&219.4&219.7&1858&\textbf{0.5}&\textbf{101.2}&\textbf{101.7}&\textbf{0.6e-4}\\

        CUTh7&8251&5.2&7557.3&7562.5&-&-&OOM&8261&8.3&212.1&220.4&\textbf{164}&1.0&276.0&277.0&1633&\textbf{0.6}&\textbf{110.9}&\textbf{111.5}&\textbf{2.5e-4}\\
        \hline
        \hline

        Avg.Spd&-&-&-&$1.0\times$&-&-&$8.4\times$&-&-&-&$25.9\times$&-&-&-&$21.8\times$&-&-&-&$\mathbf{73.6\times}$&-\\
        \hline
    \end{tabular}
    }
\end{table*}

%% file: tables/table_ablation_spmv.tex
\begin{table}[t!]
    \centering
    \caption{Ablation study to demonstrate the individual contribution of Quad-Diag format and SpMV kernels with coalesced memory access. Mixed-precision strategy is turned off and fixed FP64 is used to ensure the comparison fairness.}
    \label{tab:abl_spmv}
    \resizebox{\linewidth}{!}
    {
    \begin{tabular}{c||c|cc|cc}
        \hline
        \multirow{2}{*}{BMs}&\multirow{2}{*}{$N_{SpMV}$}&\multicolumn{2}{c|}{CSR+Irregular SpMV}&\multicolumn{2}{c}{Quad-Diag+Coalesced SpMV}\\
       
        &&$T_{SpMV}$&$T_{sol}$&$T_{SpMV}$&$T_{sol}$\\
        \hline
        \hline
        CUTh1&1334&$0.6$&$1.0$&\textbf{0.3}&\textbf{0.6}\\

        CUTh2&1675&$1.3$&$2.5$&\textbf{0.6}&\textbf{1.7}\\

        CUTh3&2078&$2.8$&$5.5$&\textbf{1.3}&\textbf{3.9}\\

        CUTh4&4018&$4.0$&$7.7$&\textbf{1.8}&\textbf{5.2}\\

        CUTh5&5918&$5.9$&$11.2$&\textbf{2.8}&\textbf{8.1}\\

        CUTh6&7408&$7.4$&$14.1$&\textbf{3.5}&\textbf{10.1}\\

        CUTh7&10232&$10.2$&$19.1$&\textbf{4.7}&\textbf{13.7}\\
        \hline
        \hline

        Avg.Spd&-&$1.0\times$&$1.0\times$&$\mathbf{2.1\times}$&$\mathbf{1.5\times}$\\
        \hline
    \end{tabular}
    }
\end{table}

%% file: tables/table_ablation_mixed.tex
\begin{table*}[t!]
    \centering
    \caption{Ablation study to demonstrate the individual contribution of the adaptive mixed-precision strategy of PCG. The solution results using FP64 are regarded as the ground truth in the calculation of $\|\bm{e}\|_{\infty}$. Fixed FP16 fails to achieve convergence.}
    \label{tab:mixed_precision}
    \begin{tabular}{c||cc|ccc|ccc||ccc}
        \hline
        \multirow{2}{*}{BMs}&\multicolumn{2}{c|}{Fixed FP64}&\multicolumn{3}{c|}{Fixed FP16}&\multicolumn{3}{c||}{Fixed FP32}&\multicolumn{3}{c}{Adaptive Mixed-Precision}\\
        &$N_{ite}$&$T_{solve}$&$N_{ite}$&$T_{solve}$&$\|\bm{e}\|_{\infty}$&$N_{ite}$&$T_{solve}$&$\|\bm{e}\|_{\infty}$&$N_{ite}$&$T_{solve}$&$\|\bm{e}\|_{\infty}$ \\
        \hline
        \hline
        
        CUTh1&\textbf{42}&0.6&-&-&$\infty$&\textbf{42}&0.3&4.0e-3&44&\textbf{0.3}&\textbf{1.4e-4} \\
        CUTh2&\textbf{53}&1.6&-&-&$\infty$&\textbf{53}&0.6&4.0e-2&56&\textbf{0.6}&\textbf{2.7e-4} \\
        CUTh3&\textbf{66}&3.7&-&-&$\infty$&\textbf{66}&1.5&0.2&71&\textbf{1.5}&\textbf{3.1e-4} \\
        CUTh4&\textbf{74}&5.2&-&-&$\infty$&\textbf{74}&2.4&8.0e-2&76&\textbf{2.3}&\textbf{2.4e-4} \\
        CUTh5&\textbf{85}&7.9&-&-&$\infty$&\textbf{85}&\textbf{4.0}&0.6&91&4.2&\textbf{2.6e-4} \\
        CUTh6&\textbf{85}&10.1&-&-&$\infty$&\textbf{85}&\textbf{5.1}&0.6&96&5.2&\textbf{2.2e-4} \\
        CUTh7&\textbf{98}&13.6&-&-&$\infty$&\textbf{98}&\textbf{6.8}&0.4&107&7.1&\textbf{2.0e-4} \\
        \hline
        \hline
        
        Avg.Spd&-&$1.0\times$&-&-&-&-&$2.2\times$&-&-&$\mathbf{2.2\times}$&-\\
        \hline
    \end{tabular}
\end{table*}

%% file: tables/table_ablation_preconditioning.tex
\begin{table*}[t!]
    \centering
    \caption{Ablation study to quantify the synergy between the Chebyshev preconditioner and our optimization suite (Quad-Diag, coalesced SpMV, and mixed-precision strategy). The optimization is toggled on/off to isolate each component's contribution.}
    \label{tab:abl_pre}
    \resizebox{\textwidth}{!}
    {
    \begin{tabular}{c||cccc|cccc||cccc}
        \hline
        \multirow{3}{*}{BMs}&\multicolumn{4}{c|}{Conjugate Gradient (CG)}&\multicolumn{4}{c||}{Jacobi-Preconditioned CG}&\multicolumn{4}{c}{Chebyshev-Preconditioned CG} \\
        &\multicolumn{2}{c}{Optimization Off}&\multicolumn{2}{c|}{Optimization On}&\multicolumn{2}{c}{Optimization Off}&\multicolumn{2}{c|}{Optimization On}&\multicolumn{2}{c}{Optimization Off}&\multicolumn{2}{c}{Optimization On} \\
        &$N_{ite}$&$T_{tot}$&$N_{ite}$&$T_{tot}$&$N_{ite}$&$T_{tot}$&$N_{ite}$&$T_{tot}$&$N_{ite}$&$T_{tot}$&$N_{ite}$&$T_{tot}$ \\
        \hline
        \hline
        CUTh1&1192&1.3&1192&1.2&647&0.9&648&0.8&\textbf{44}&1.1&\textbf{44}&\textbf{0.5}\\
        CUTh2&1523&2.6&1524&2.4&823&1.6&823&1.5&\textbf{54}&2.4&56&\textbf{0.8}\\
        CUTh3&1939&5.8&1939&5.1&1034&3.6&1034&3.4&\textbf{67}&5.4&71&\textbf{1.8}\\
        CUTh4&2187&8.2&2187&7.1&1150&5.0&1150&4.5&\textbf{75}&7.7&76&\textbf{2.7}\\
        CUTh5&2493&12.3&2494&10.8&1427&8.2&1427&7.4&\textbf{86}&11.2&91&\textbf{4.7}\\
        CUTh6&2710&16.6&2712&13.4&1428&9.5&1428&8.7&\textbf{90}&13.7&96&\textbf{5.7}\\
        CUTh7&3024&21.5&3026&18.3&1593&13.0&1593&11.8&\textbf{99}&19.3&107&\textbf{7.7}\\
        \hline
        \hline

        Avg.Spd&-&$1.0\times$&-&$1.1\times$&-&$1.6\times$&-&$1.7\times$&-&$1.1\times$&-&$\mathbf{2.9\times}$\\
        \hline
    \end{tabular}
    }
\end{table*}

%% file: chapters/6_conclusion.tex
\section{Conclusion and Future Work}
\label{sec:6}
\textbf{Conclusion:} This paper has presented \textbf{CUTh-Solver}, an open-source GPU-accelerated sparse solver framework tailored for high-resolution 3D IC thermal analysis. We have proposed a Quad-Diag storage format with coalesced SpMV kernels, a highly parallel Chebyshev preconditioning strategy, and an adaptive mixed-precision scheme. Extensive evaluations demonstrate that our proposed CUTh-Solver achieves significant speedups over the latest GPU-accelerated COMSOL Multiphysics 6.4 and a wide range of NVIDIA's native sparse libraries.

\textbf{Future Work:} CUTh-Solver targets 7-diagonal banded SPD linear systems, which arise from the popular 7-point finite difference/volume discretization of 3D IC thermal simulation. The main limitation of CUTh-Solver is that Quad-Diag does not support unstructured meshes (e.g., tetrahedral cells of FEM), where the diagonal regularity is lost. Extending the framework to hybrid DIA–ELL or DIA–COO for unstructured discretizations is left for future work.

%% file: chapters/7_acknowledgment.tex
\section{Acknowledgment}
\label{sec:7}
AI assistants were used for language polishing and partial code generation under the authors' supervision. All research decisions and contributions were made by the authors.